\documentclass[]{aa}
\usepackage{graphicx}
\usepackage{amssymb}
\begin{document}
\input{psfig.sty}
 	
\title{Extragalactic Globular Clusters in the Near Infrared:
}
\subtitle{IV.~Quantifying the Age Structure using Monte-Carlo Simulations}

\author {Maren Hempel
        \and
        Markus Kissler-Patig
	}
\offprints {Maren Hempel} 

\institute{European Southern Observatory, Karl-Schwarzschild-Str.~2,
85748 Garching, Germany \\ \email{mhempel@eso.org, mkissler@eso.org}}

\date{Received 17 October 2003; accepted 23 February 2004 }

\abstract{In previous papers of the series, we used a combination of
optical and near-infrared colours to derive constrains on the relative
age structure in globular cluster systems. Here, we present the
details, strength and limitations of our method based on Monte-Carlo
simulations of colour-colour diagrams and cumulative age
distributions. The simulations are based on general informations about
the globular cluster systems (e.g. colour-ranges, the number ratios
between sub-populations) and the different single stellar population
models (SSP's) which are used to derive relative ages. For both the
modeled systems and the observed globular cluster systems we derive
the cumulative age distribution and introduce two parameters to define
it, the so-called 50$\%$ age and the result of the reduced
$\chi$$^2$test of the comparison between models and observations. The
method was tested successfully on several systems and allowed to reveal
significant intermediate age populations in two of them.

}

\authorrunning{Hempel, Kissler-Patig}
\titlerunning{Age distribution in Globular Cluster Systems {\tt
IV}}
\maketitle

\section{Introduction}
\label{s:intro}

Globular cluster systems have been extensively used in the last decade
to interpret the formation and evolution histories of early-type
galaxies (see for example Geisler, Grebel \& Minniti (eds.) 2002 and
Kissler-Patig (ed.) 2003). A key question remains the origin of the
multiple globular cluster sub-populations observed within most (if not
all) individual galaxies. The metal-poor, halo sub-populations appear
very uniform and are most probably linked to star cluster formation in
the early structures of the universe (e.g. \cite{ash94},
\cite{burgarella01}). However, the metal-rich sub-populations are less
likely to be homogeneous in age and/or metallicity, as nicely seen
e.g. in the sample of Kundu et al. (2001a,b) and might contain
information on the more recent epochs of star formation in the
galaxies. While certainly a large fraction of metal-rich globular
clusters investigated so far is old (e.g. \cite{puzia99}), the existence of
large intermediate-age globular cluster populations has not been excluded yet and was
actually shown to exist in at least some galaxies
(e.g.~\cite{goudfrooij01}, \cite{puzia02},
\cite{hempel03}).

This latter fact is closely related to one of the most important open
questions in galaxy formation and evolution, namely whether galaxies
formed the vast majority of their stars at early epochs, or whether a
significant fraction of the stars originated from more recent events,
such as dissipational mergers (e.g.~\cite{kennicutt98b}, \cite{renzini99a},
\cite{renzini99b}, \cite{schweizer00} and references therein).

The study of globular clusters in order to determine the major star
formation epochs presents several advantages over the study of the
diffuse galaxy light. Indeed-~several age populations can be hidden in
the diffuse light of the host galaxy (e.g.~\cite{larsen03}). Globular
clusters on the other side represent single stellar populations, their
stars sharing the same metallicity and the same age. The existence of
sub-populations of globular clusters gives henceforth a stronger
indication for different star formation epochs. Since the prediction
and first discovery of multiple globular cluster sub-populations in
early-type galaxies by Ashman \& Zepf (1992, see also Zepf \& Ashman (1993))
their colour distributions have been investigated extensively. Various
studies, e.g.~by Gebhardt \& Kissler-Patig (1999) and Kundu \& Whitmore
(2001a,b) have shown that bimodal colour distributions in elliptical
galaxies are common. The exact interpretation, however, is hindered by
the age-metallicity degeneracy of broad band colours
(\cite{worthey94}).

Our group uses combined optical- and near infrared photometry 
(Kissler-Patig et al.~2002, Puzia et al.~2002, \cite{hempel03}, 
hereafter cited as Paper I,~II,~III ) to overcome this obstacle and
partly lift this degeneracy. In Paper III, we showed how, using Single Stellar 
Population (SSP) models and the corresponding model isochrones, 
(e.g.~by Bruzual \& Charlot 2000, Vazdekis 1999 and Maraston 2001) 
photometric studies of globular cluster systems can 
be used to derive their age distributions. 

In this paper we present the details of our method, investigate the capability
of photometric studies to detect globular cluster sub-populations of
different ages within the metal-rich sub-population and the accuracy
with which we can derive their relative ages. 

The paper is organized as follows: in Section \ref{s:modelling} we
will describe our method of simulating colour-colour diagrams for
globular cluster systems. Section 3 shows how these artificial
distributions are compared to the observed data via their cumulative
age distributions. Caveats and uncertainties of the method are
presented in Sect.~4, first results and our conclusions are given in
Sect.~5 and 6.

\section{Modeling of colour distributions}
\label{s:modelling}
\subsection{Overview of the method}

\begin{figure}[h]
\centering
\includegraphics[bb=50 150 580 710, width=8cm]{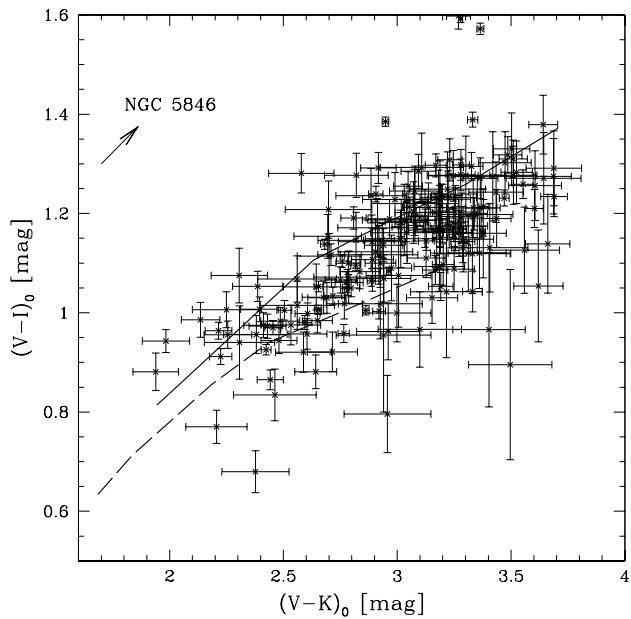}
\caption{Colour--colour diagram for the NGC~5846 globular cluster system.
 The solid and dashed line mark the 15 Gyr and 2 Gyr isochrone respectively following the
 Bruzual \& Charlot SSP model (Bruzual \& Charlot 2000).}
\label{f:n5846colcol}
\end{figure}

The basic idea of our approach is best described by a question: How
would a colour-colour diagram of a globular cluster system look like if
it was indeed {\it{formed in two or more distinguishable star
formation events}}? We decided to model different cases and compare these 
to our data (Fig. \ref{f:n5846colcol}, see also Paper III). 

Originally, our ambition was a direct detection of the different
globular cluster sub-populations in a given galaxy in our data
(colour- colour diagrams) by comparing observed colour distributions
with simulations following SSP models. This, however, turned out to be
too difficult with non-converging solutions. Indeed for the direct
comparison of observed and simulated colour-colour diagrams we applied
a two-dimensional Kolmogorov-Smirnov test (\cite{press92}). However we
had to consider a number of parameters which are included in our
observations but difficult to include in the simulations,
e.g. completeness limits, photometric error with high enough accuracy
to still disentangle in 2 dimensions the parameters of interest (age,
number ratios of the different populations. In Sect.\ref{s:age} we
describe our alternative, 1- dimensional approach via the cumulative
age distribution. \\

At this stage we do not address the question of what actually ignited
the cluster formation and are working with the final observed colour
distribution only. Also, at this point, we stay with the assumption of
two major star formation events and a first generation age of 15 Gyr
out of convenience. The latter does not play a major role since we are
interested in relative ages only. Further, if we take into account the
age uncertainty of the model isochrone in the age range between 10 and
15 Gyr our assumption is still reasonable with respect to the new WMAP
results giving a maximum age of the universe of 13.7 Gyr $\pm$ 0.2 Gyr
(e.g. \cite{bennett03}). We will also present results for
NGC~5846 assuming the old population of age 10 Gyr in Sect. 5.1.. \\

The basic principle is thus to input a number of properties for the two
sub-populations to model (see next section), use a Monte-Carlo approach
to model the distribution of the first colour, determine the second colour with
the help of SSP's, construct a cumulative age distribution from this
artificial two-colour diagram and compare the latter to its observed
counter-part. These steps are repeated for various input variables that
scan the two parameter space (size of the second populations, age of
the second population).

To illustrate the method, we use data presented in Paper III: we model
colour--colour diagrams, $(V-I)$~$vs.$~$(V-K)$, for a template
globular cluster systems corresponding to the one of NGC~5846. This is
a favorable example given the relatively large number of objects with
VIK colours (188 globular clusters for the complete sample, when no
limits for photometric errors, colour or limiting magnitudes are
applied). In Sect.~4, we will also illustrate the caveats for smaller
samples and the possibility of contamination.

An important note is that we are interested in the age structure of
the metal-rich population only. We assume the {\it blue}
(2.0$\leq$$(V-K)$$\leq$~2.7) population to consist of {\it old}
objects only (in NGC~5846 43 cluster were assigned to the metal-poor
population), and model exclusively the red (2.7$\leq$$(V-K)$$\leq$3.8)
colour range.  We generate for our example case in total 120 metal-
rich objects, split into an {\it old} and {\it intermediate age}
population.


\subsection{Input parameters to the simulation}

Several input parameters get fixed with respect to the observed
distribution.
\begin{itemize}
\item  {\bf{the total number of metal-rich objects}}, in our example we
have 120 metal-rich globular cluster candidates and simulate
individual colour distributions with that number of objects
\item  {\bf{the primary colour range to be modeled}}, in our template
$(V-K)$ is used as primary colour in the range $2.7<(V-K)<3.8$. The
lower limit is driven by the observed gap between blue and red
sub-populations, the upper limit encloses metallicities of up to twice
solar at the highest ages, but excludes redder contaminating galaxies 
\item {\bf the SSP model} to be used in order to associate, for a given age,
the secondary colour (in our case $(V-I)$) to each primary colour point
\item {\bf the observed error distribution in the primary colour} which
is used to randomly ``smear'' the primary colour after calculating the
secondary colour
\item { \bf the observed error distribution in the secondary colour} which
is used to randomly ``smear'' the secondary colour
\end{itemize}

For each study of a particular data set these fixed parameters can be
chosen such as to mimic closely the observed data. In addition to
these fixed parameters two variable input parameters are of interest:
how many young clusters are present, and what is their age?

\begin{figure}[h]
\centering
\includegraphics[bb=50 150 580 710, width=8cm]{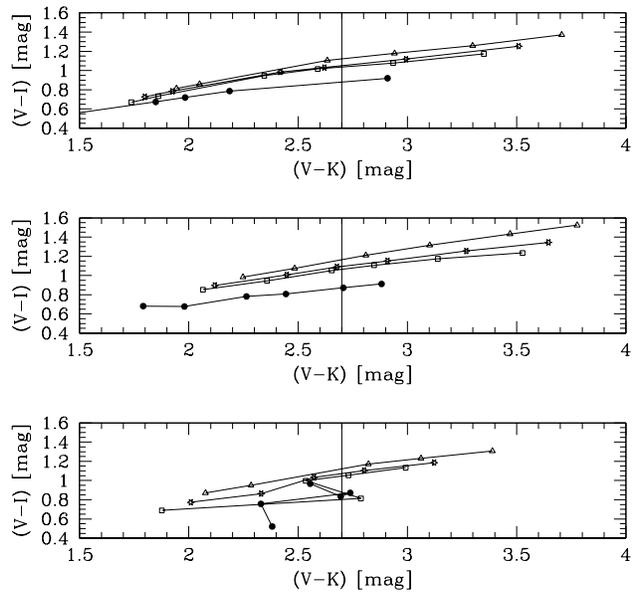}
\caption{Comparison between the  SSP model isochrones given by
Bruzual \& Charlot (2000) (upper panel), by Vazdekis (1999) (middle
panel) and Maraston (2001) (lower panel). Different symbols mark the 1
Gyr (solid circles), 3 Gyr (open squares), 5 Gyr (open stars) and 15
Gyr (open triangles) isochrones. The solid line at $(V-K)$=2.7 marks
boundary between {\bf blue} and {\bf red} objects (see section \ref{s:monte}). }
\label{f:SSPmodel}
\end{figure}


\subsection{Variables of the simulation}
\label{s:variables}
Our scientific goal is to investigate {\it i)} whether a second,
intermediate-age population is present in the metal-rich sub-population;
and if so {\it ii)} what is the most likely age of the population and
{\it iii)} what is its importance in number of clusters relative to the
old metal-rich population. Thus, we explore the parameter space span by
the two latter properties. To do so, each set of simulations 
is performed for a pair of these variables: 

\begin{itemize}
\item { \bf The relative number ratio of the old with respect to the 
intermediate age sub-population}.  We vary the ratio of both populations
between a populations composed of 100\% 15 Gyr old clusters 
(no younger sub-population present) to 100\%  young objects, in 10\% 
increments. 
\item { \bf The age of the second population} for which we modeled the
cases of 1, 2, 3, 5, 7, and 10 Gyr.
\end{itemize}

Thus, simulations are preformed for each of the 66
possible pairs of variables. This leads to 66 master models which get
compared to the observed data.


\subsection{Monte-Carlo simulation to get a set of primary colours}
\label{s:monte}

For each individual simulation, we start by creating an artificial
globular cluster samples in our primary colour. We populated the two
$(V-K)$ colour intervals assigned to a {\bf ``blue''} and a
{\bf ``red''} sub-population with an fixed number of objects. Hereby
we assume a random distribution within this range. A second run of
simulations using a Gaussian colour distribution in both populations
did not lead to significantly different results than the once obtained
for random distributed objects. Given the lack of physical support for
Gaussian colour distributions, we stay with the simplest case: a pure
random distribution.


\subsection{Determination of the second colour and observational errors}
\label{s:second}
The importance of using different SSP models, e.g. by Bruzual \&
Charlot (2000, hereafter named BC00), Vazdekis (1999, hereafter named
VA99) or Maraston et al. (2001, hereafter named MA01) for our modeling
becomes clear, if we compare the corresponding isochrones. In Fig.
\ref{f:SSPmodel} the isochrones for a 1,~3,~5,~10 and 15 Gyr old SSP
are shown following the model by Bruzual \& Charlot (2000), Vazdekis
(1999) and Maraston (2001). Especially in the red $(V-K)$
colour range we find a significant discrepancy for the corresponding
$(V-I)$, resulting in a model dependency of the derived
ages. Therefore we will apply different models in the age dating.

\begin{figure}[h]
\centering
\includegraphics[bb=50 470 580 710, width=8cm]{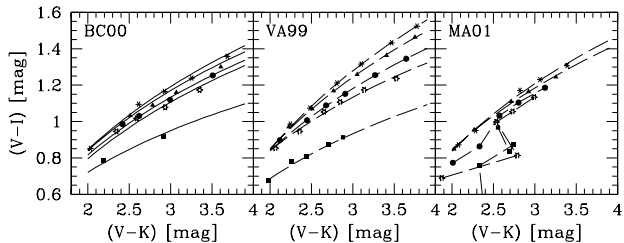}
\caption{Least square fit of model isochrones given by Bruzual \&
Charlot (2000)(left), Vazdekis (1999) (center) and Maraston (2001)
(right). Various isochrones are marked with solid squares (1Gyr), open
stars (3 Gyr), solid circles (5 Gyr), solid triangles (10 Gyr) and
asterisks (15 Gyr).}
\label{f:com_model}
\end{figure}

The next step in the simulation is the parametrisation of the model
isochrones. A least square fit of $(V-I)$ as a logarithmic function of
$(V-K)$ in the form $(V-I)$=A$\cdot$ln$(V-K)$+B, as shown in Fig.
\ref{f:com_model}, gives us a set of parameter A and B for all SSP
models. Unfortunately the Maraston isochrones for SSP's in the lower
age range (less than 5 Gyr) allow no simple parametrisation. We
therefore decided to postpone the work with this specific model. The
relations were used to calculate the {\bf secondary} colour $(V-I)$
for each assumed age population. We generated colour-colour diagrams
for globular cluster systems consisting of an old (15 Gyr) population
and a se\-cond one, as explained in Sect.
\ref{s:variables}. Each simulation will therefore create the $(V-I)$
$vs.$$(V-K)$ colour-colour diagram for a mixed population of globular
clusters. Finally both colours $(V-K)$ and $(V-I)$ were smeared with
up to 3$\sigma$ photometric error, taken from the NGC~5846 observed
error catalog. In a last step, an artificial catalog with the two
colours and their 1$\sigma$ error is created. The final outcome of
this process is the colour-colour distribution
$(V-I)$$\pm$~$\Delta$$(V-I)$~$vs.$$(V-K)$$\pm$~$\Delta$$(V-K)$ for six
age combinations and eleven number ratios. We generated 1000 models
for each set of model parameters. The final product, the cumulative
age distribution (see section \ref{s:cumulative}) to be compared with
the observations, was determined as the statistical mean of the
distribution of these 1000 models.

As we can see in Fig. \ref{f:com_model}, the isochrones for various
SSP models (e.g. Bruzual \& Charlot, Vazdekis and Maraston) differ
strongly. Nevertheless there is no quality argument
(\cite{vazdekis96}) behind our choice of BC00 and VA99. Since Bruzual
\& Charlot provide models for the largest variety of ages and
metallicities we started our modeling program using parametrised BC00
isochrones (Fig. \ref{f:com_model}, left panel). VA99 has been added
later to test our method for model dependence. The latter will be
discussed in section \ref{s:results}. Both models assume a
Salpeter IMF.

\section{Quantifying the age structure}
\label{s:age}
\subsection{Cumulative age distribution}
\label{s:cumulative}

\begin{figure}[h]
\centering
\includegraphics[bb=50 150 580 710, width=6cm]{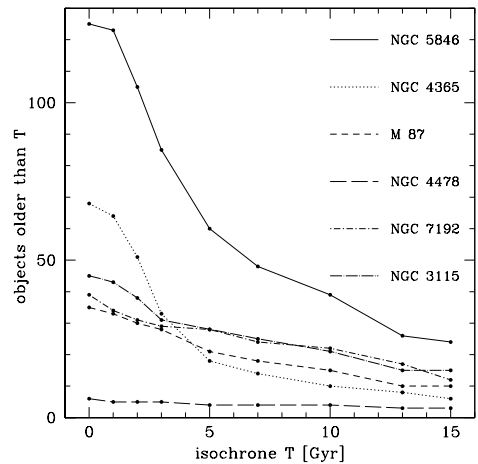}
\caption{Cumulative age distribution in the globular cluster
systems of NGC~5846, NGC~4365, NGC~7192, NGC~3115, NGC~4478 and M87.}
\label{f:absolutage}
\end{figure}

Photometric errors, which might become rather substantial in combined
optical and near-infrared photometry, contribute largely to the
uncertainty in age determination. Comparing our limiting photometric
error of 0.15 mag and the colour difference in $(V-I)$ for SSP's which
differ in age by at least 1 Gyr (see Fig. \ref{f:com_model}), we
conclude that not only absolute ages are out of reach but also age
resolution more accurate than several Gyr. Nevertheless- our major
goal is the detection of cluster sub-populations in order to give
evidence for various major star formation events during the evolution
of the galaxy. Since we assume this events to be caused by merging or
accretion, likely but not very frequent events, age differences larger
than 5 Gyr are to be expected. The cumulative age distribution (see
Fig. \ref{f:absolutage}) which we derive is only based on the object
density in the colour-colour diagram with respect to the isochrones
and is less sensitive to photometric errors compared to a direct age
determination. Since the result will refer to the globular cluster
system the size of the sample becomes important, as we will show in a
later section (see Sect. \ref{s:stability}).

The derivation of the cumulative age distribution is a straight
forward procedure and is identical for observed and simulated
samples. Each object was assigned to an age older than X if its
$(V-I)$ colour was found to be redder than for the isochrone of that
age. The model isochrones for different ages at low $(V-K)$ values are
hard to distinguish and the $(V-I)$ colour of a specific isochrone
depends only weakly on $(V-K)$ in the red-~colour regime. Therefore we
set a lower colour limit (see also Paper III) for $(V-K)$ of
$\leq$2.6, corresponding roughly to a metallicity of
[Fe/H]$\sim$-0.7dex, slightly above the split between metal poor and
metal rich clusters in galaxies. 

\begin{figure*}[!t]
\begin{flushleft}{
\includegraphics[width=5.5cm]{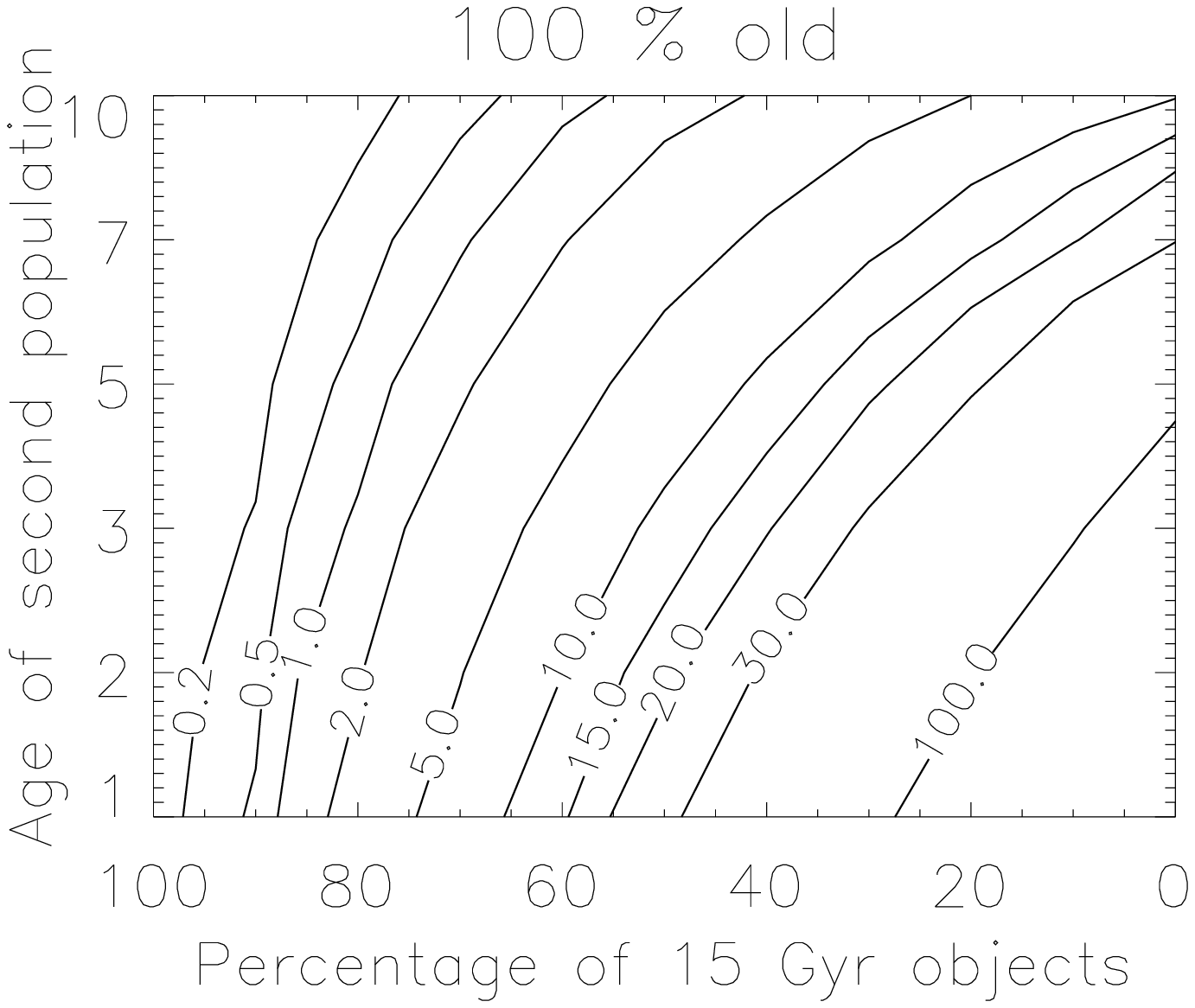}
\includegraphics[width=5.5cm]{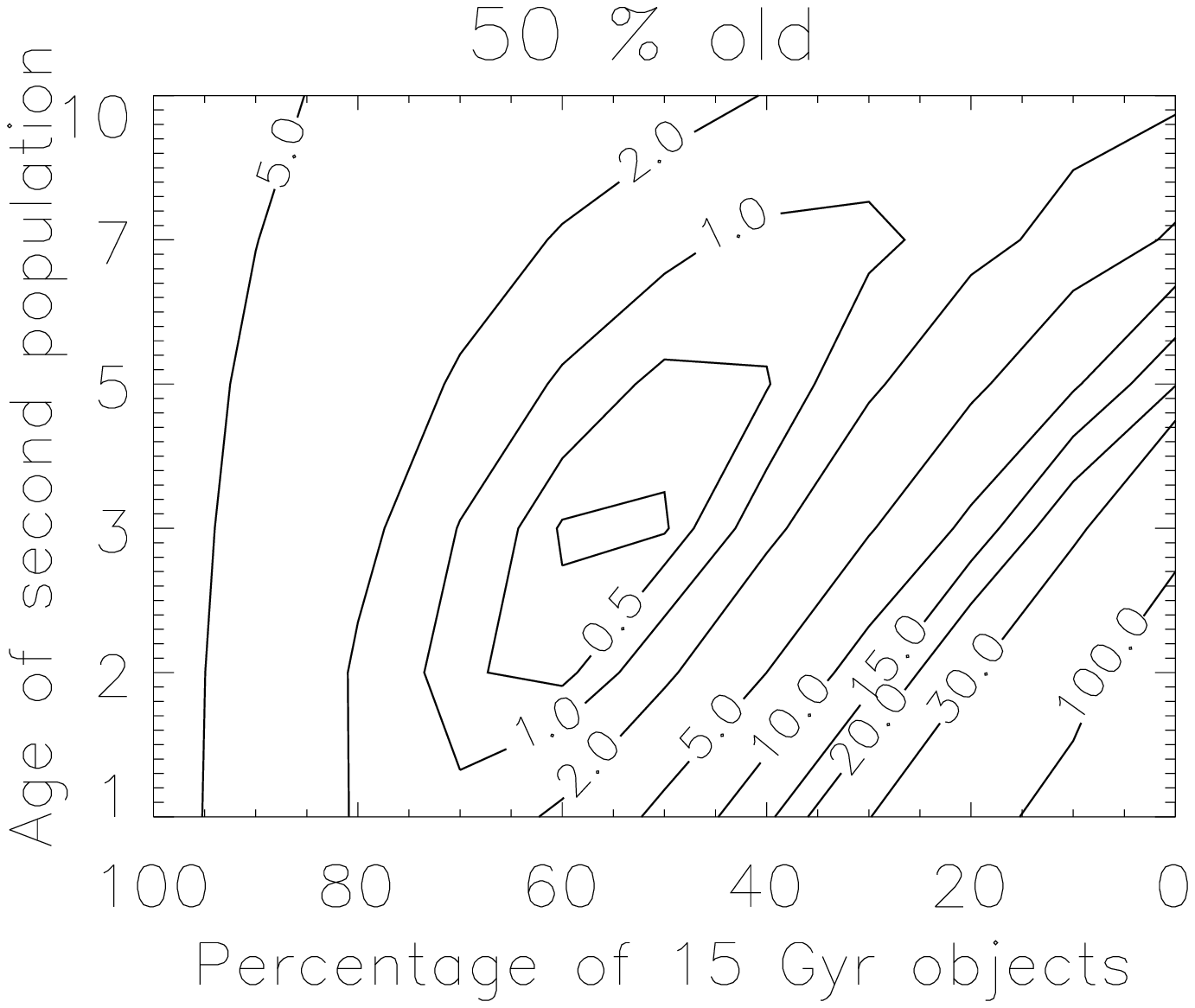}
\includegraphics[width=5.5cm]{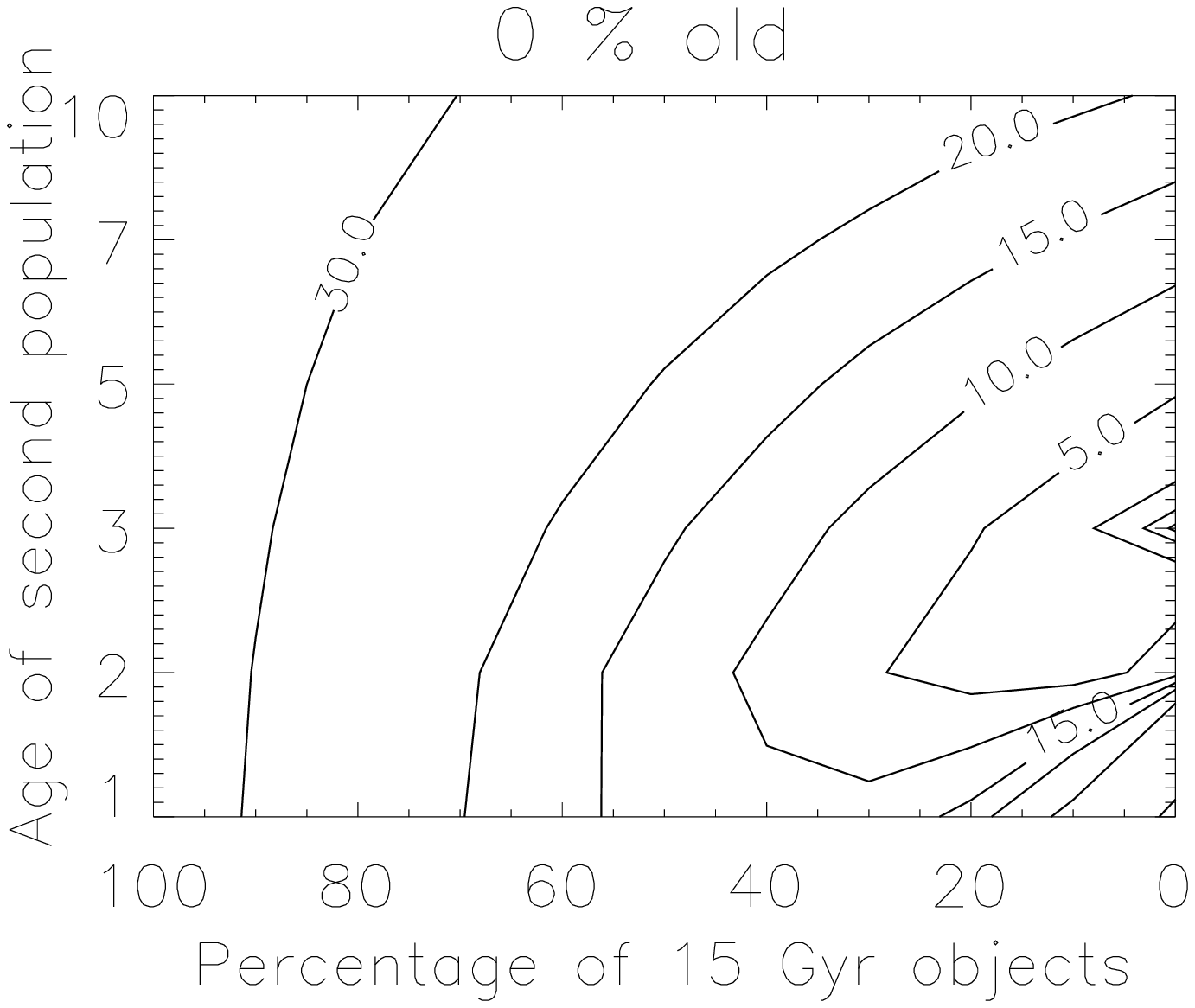}
\includegraphics[width=5.5cm]{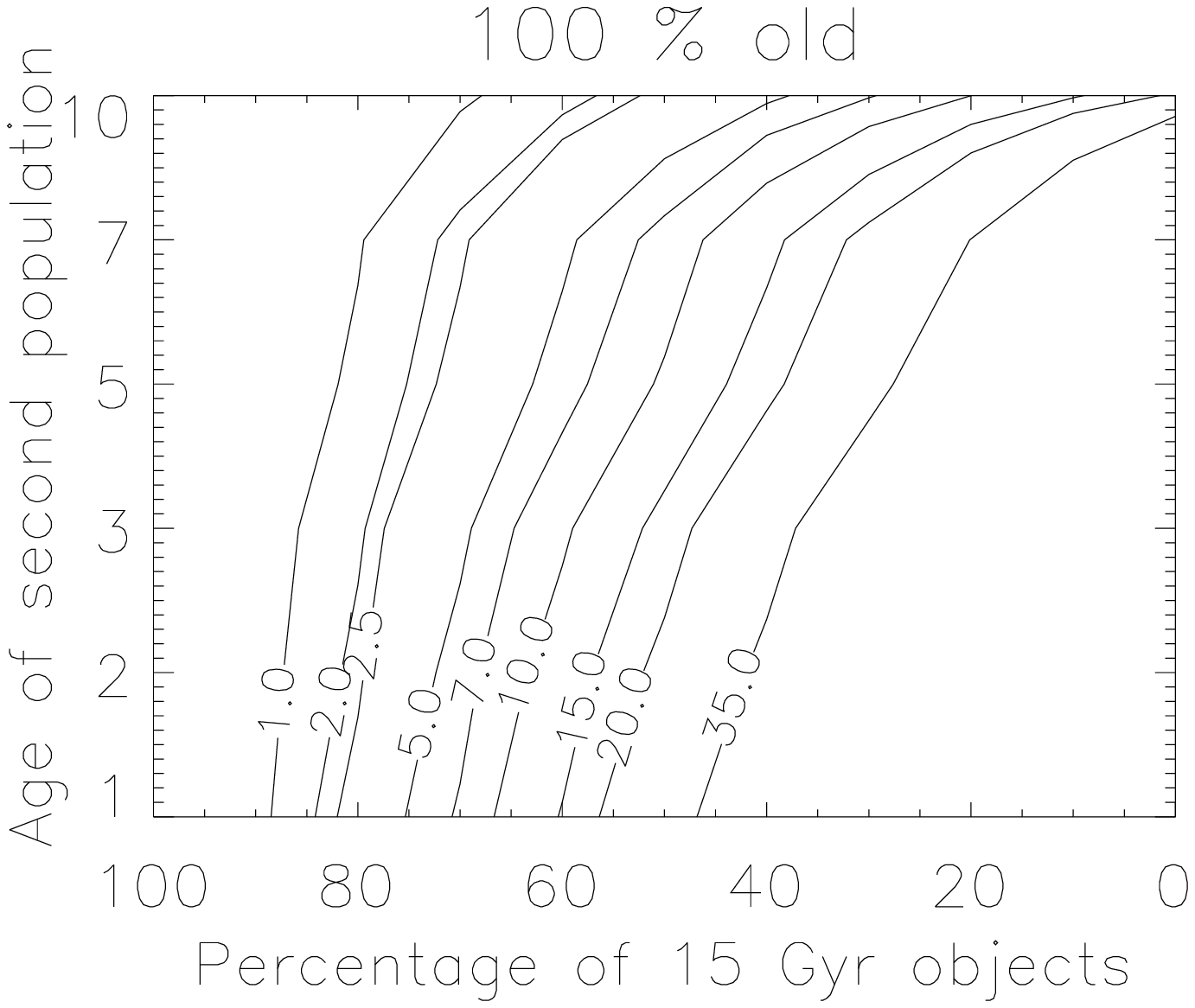}
\includegraphics[width=5.5cm]{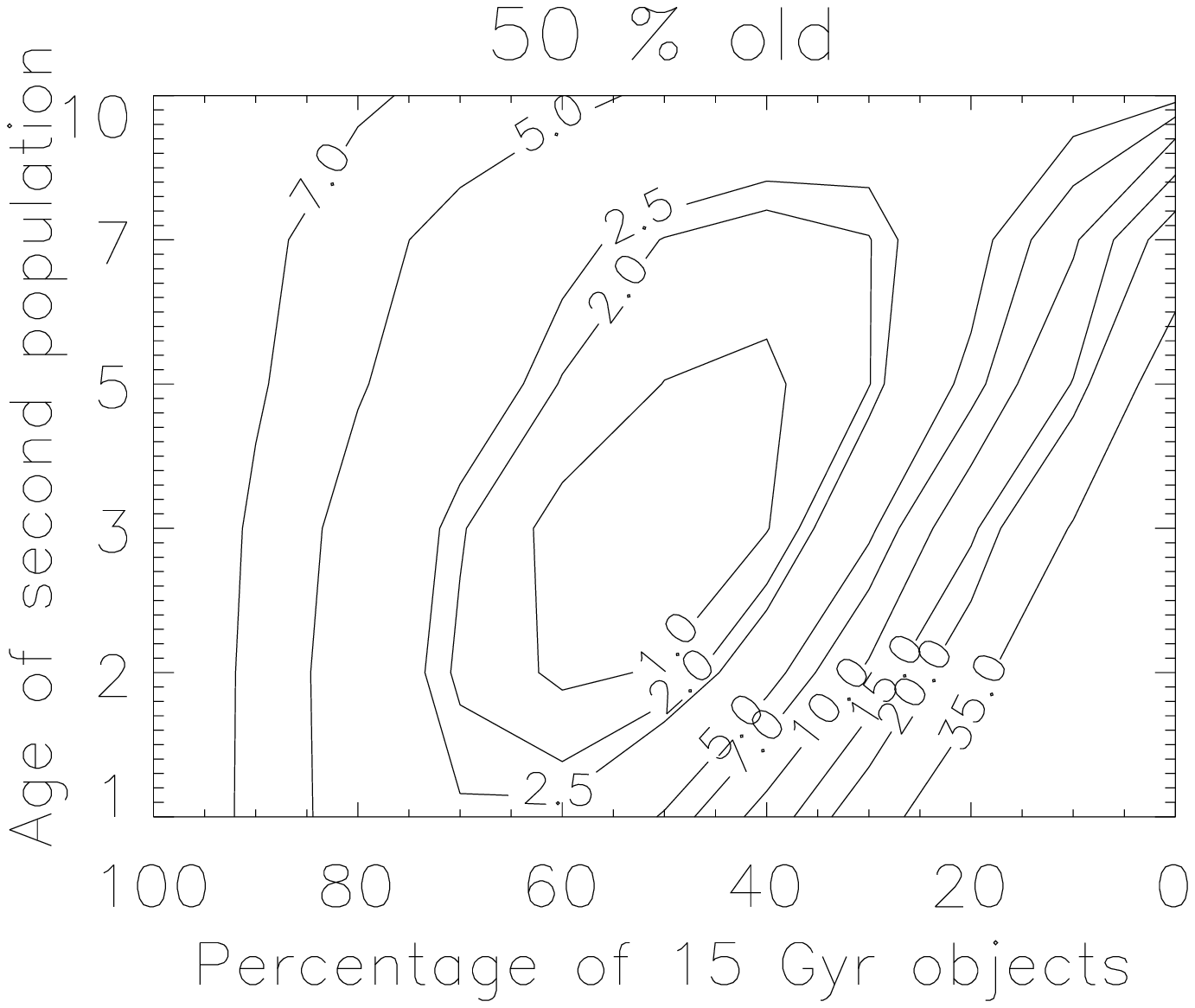}
\includegraphics[width=5.5cm]{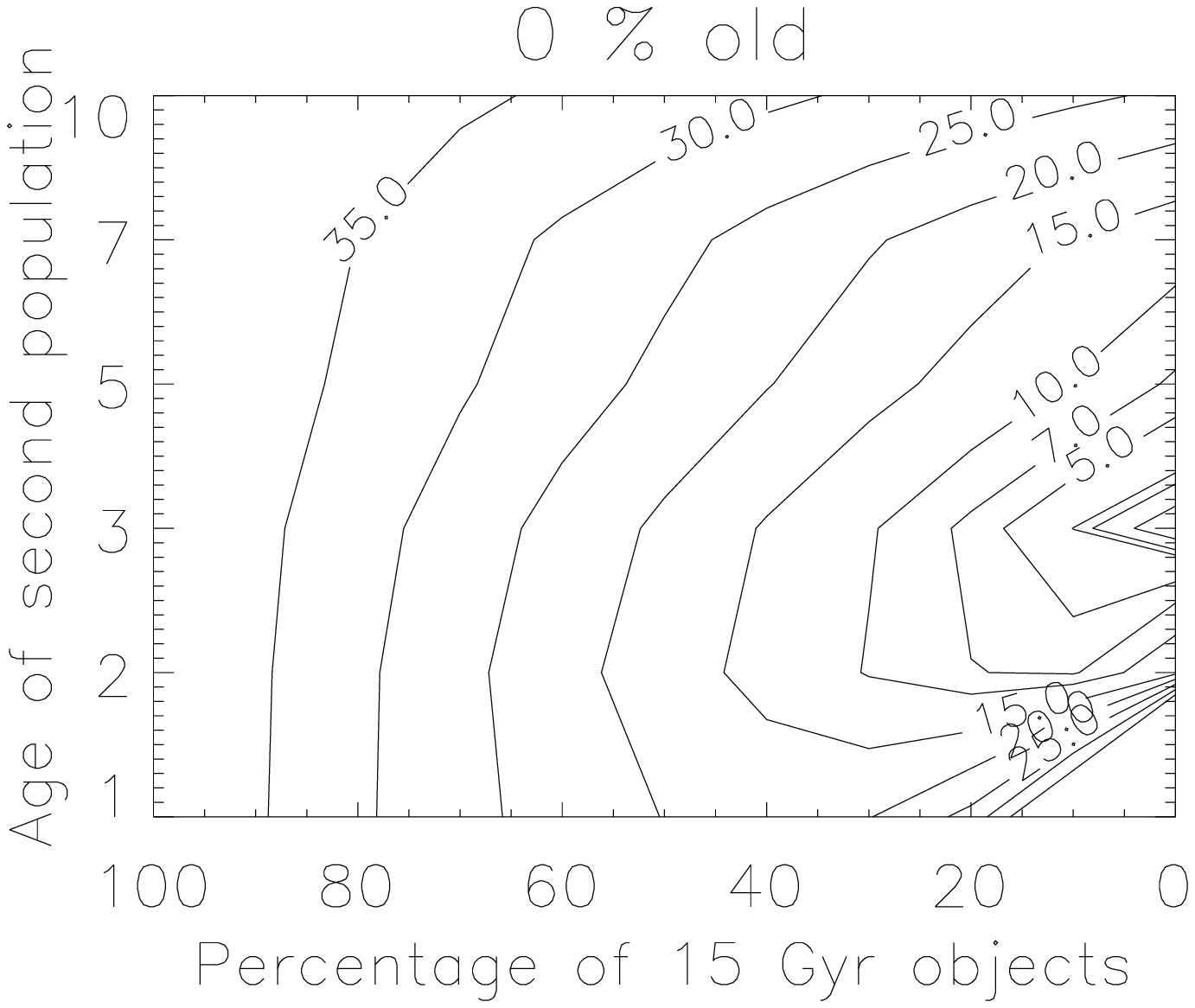}
\caption{Self-consistency test: age distributions of individual model systems
compared to the complete model set. The result of the $\chi$$^2$ test
are shown as $\sigma$- contours. For this test we use the model of a
purely 15 Gyr old population (left), a 15 Gyr + 3 Gyr mixture (middle)
and a purely 3 Gyr old population (right). As expected the result for
the purely old population is much more uncertain than for the other
two, allowing as best fitting model also mixed populations. The upper
plots show the result for the BC00 model and the lower plots the VA99
models. The different spread of the isochrones (see Fig.
\ref{f:SSPmodel}) results in different uncertainties.}
\label{f:comparemodel}}
\end{flushleft}
\end{figure*}

The upper limit was set to $(V-K)$
$\geq$ 3.6, which agrees with the upper $(V-K)$ limit in our NGC~5846
sample. The age distributions are normalised with respect to the total
number of objects within 2.6$\leq$$(V-K)$$\leq$3.6. The notation 0 Gyr
in the cumulative age distribution refers to the total number of
objects (per definition all objects are 'older' than 0 Gyr).

\subsection{The comparison with observed data: $\chi^2$ test}

The relative size of globular cluster sub-populations provides some
constrains for the galaxy formation scenario. In Ashman and Zepf
(1992) the efficiencies of the globular cluster formation in
dependency of the environment (cluster galaxies or 'normal' elliptical
galaxies) and the amount of available gas are compared and found to
differ significantly. Therefore we are interested in the ratio between
possible globular cluster sub-populations. Since the globular cluster
samples are limited to the inner most region of the globular cluster
systems this result will only answer the question whether there is a
second population and to which extend we can constrain its age and
relative size compared to an old population.

To quantify the age structure we compare the cumulative age
distributions of the observed and modeled data sets. By using a
reduced $\chi$$^2$ test we find the best fitting model for the
observed system.

Due to the age uncertainty of the models, the 3$\sigma$ scatter in
colour allowed in the model, and to the photometric error in the
observations, we expect a certain degeneracy in the age/ratio
combinations. Nevertheless we are able to constrain the age as well as
the ratio between the two populations. This was tested by comparing
the age distribution for three selected models with the complete model
set. The $\chi$$^2$ test should pinpoint the model which was chosen in
the first place. In Fig. \ref{f:comparemodel} we show the result for
the $\chi$$^2$ test for models of a 100$\%$~15 Gyr, a 50$\%$~15 Gyr +
50$\%$~3 Gyr and a 100 $\%$~3 Gyr population respectively. As we can
see in the plots the $\chi$$^2$ tests return the model parameters we
chose as the best fit, including an uncertainty in age and number
ratio. Due to the different $(V-I)$ colour range for BC00 (upper
panel) and VA99 (lower panel) SSP isochrones the result of the
$\chi$$^2$ test differ slightly.

\subsection{Comparison of different SSP models}
\label{s:compare}
Before we start comparing our results based on the SSP models by
Bruzual \& Charlot (\cite{charlot91}, \cite{bruzual93}) and by
Vaz\-dekis (\cite{vazdekis96}) we want to make the point that our
choice is by no means to interpret as judgment on the various
models. As Vazdekis et al. (1996) pointed out there is, despite the
different modeling procedure, little difference in the quality of the
various models, which mostly differ in the covered range of
metallicity and age and in accuracy of the physical input
parameter. The SSP models by Bruzual \& Charlot are based on the
Padova library from 1994. The newer version, based on the Padova 2000
library, is used in the latest SSP model by Bruzual \& Charlot, but
not recommended by the authors (see \cite{bruzual03}). Details can be
found in Bruzual \& Charlot (2003). For more information about the
VA99 model we refer to Vazdekis et al. (1996). The essential
difference to the BC00 models being the application of empirical
stellar libraries whereas Bruzual and Charlot make use of the
libraries by Lejeune (e.g.\cite{lejeune97}). The later are based on
the stellar atmospheres derived by Kurucz (e.g.\cite{kurucz92}). We
would like to stress again that we do NOT attempt to judge the quality
of different models but use them to discuss our method with respect to
model uncertainties. Our choice of the BC00 and VA99 model is
therefore highly subjective and mostly driven by the typical error is
introduced by parametrising the model isochrones for calculating the
secondary colour (as explained in~\ref{s:monte}). In the future, we
intend to replace the parametrised by linear interpolated secondary
colors in order to reduce this error. In the discussion of our results
(see Sect. 5.1, Fig. \ref{f:shift0}) we will present first results
based on this modification.

The smaller spread in $(V-I)$ for BC00 results in a finer scanning of
the colour-colour diagram and gives therefore better results in the
$\chi$$^2$ test. For larger sample sizes (observed and simulated
systems) this effect will become more and more negligible. We expect
the strongest effect for the comparison of small observed systems
(e.g. NGC~4478) with the relatively large (120 clusters) model systems
(see Sect. 5).

\begin{figure}[!]
\includegraphics[width=8cm]{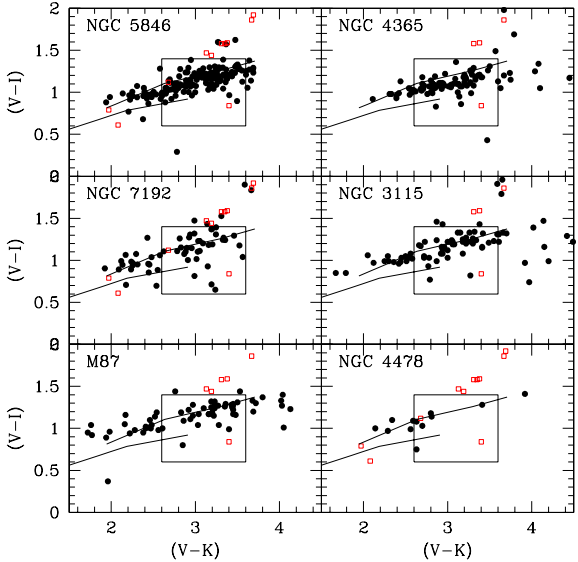}
\caption{Comparison of  colour -colour diagrams
for the globular cluster systems and the HDF-S. Globular clusters are
marked by filled circles and HDF-S objects by open squares. The solid
lines mark the 1 Gyr and 15 Gyr isochrone respectively, as given by Bruzual \&
Charlot (2000). The box marks the colour range used for determining the
cumulative age distribution.}
\label{f:compare}
\end{figure}


\section{Potential caveats}

\subsection{Contamination with background objects}
\label{s:contamination}

Using colour distributions with respect to model isochrones is
strongly affected not only by photometric errors but also by
false identification of objects. Within the red colour range,
as it has been defined in section \ref{s:cumulative}, the
contamination of our globular cluster sample by background galaxies
and/or foreground stars has to be considered. In the red $(V-K)$
colour range these objects are most likely unresolved background
galaxies.  Assuming their random distribution on the sky we use the
Hubble-Deep-Field South, available from the archive of
ST-ECF\footnote{www.stecf.org/hstprogrammes/ISAAC} as a representative
background sample. We are well aware that a random distribution of
background objects is only the case in a first approximation
(\cite{roche93}, \cite{infante95}, \cite{maller03}).

\begin{figure}[!]
\centering
\includegraphics[bb=50 430 580 710,width=8cm]{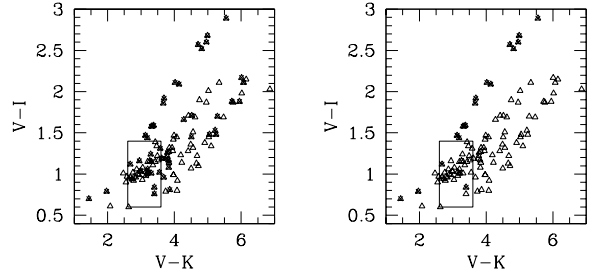}
\caption{Colour- colour diagram for the HDF-S sample (K$>$21.5) before
(triangles) and after (cross) applying shape parameter for
selection. For selection the ellipticity ($<0.2$, left panel) and the
star/galaxy classifier ($>0.8$, right panel) have been used. For final
correction we choose a combination of photometric error, limiting
magnitude and classifier. As in Fig.6 the box indicates the colour
limits.}
\label{f:colcol_hdf}
\end{figure}

In Figure \ref{f:compare} we compare the colour-colour diagrams for
observed systems, taken from Paper I, II and III, with the background
sample, to which the same selection criteria as to the cluster sample
have been applied, i.e. limits for the photometric error, the limiting
K- magnitude and the colour cuts for $(V-K)$ and $(V-I)$ (marked by
the box). Additionally the HDF objects have been selected using the
SExtractor star/galaxy classifier as shape parameter. For comparison,
the colour-colour diagram for the HDF-S sample is given in Fig.
\ref{f:colcol_hdf} using only error cuts and limiting magnitude
(symbol: $\bigtriangleup$) or including ellipticity
(left panel) and classifier (right
panel) as well. The latter was applied to our observed data sets.

\begin{figure}[]
\centering
\includegraphics[bb=50 440 580 710, width=8cm]{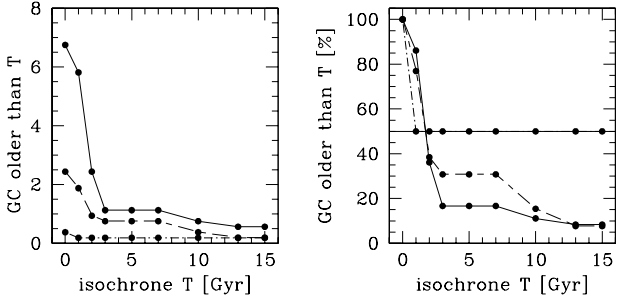}
\caption{Cumulative age distribution in the HDF-S sample (BC00)
normalised to 1 arcmin$^{2}$ (left: absolute, right: relative). The solid
line marks the result if only a error cut of 0.15 mag for all filters
is applied. Applying an additional ellipticity limit ($<$0.2)
results in a age distribution indicated by the dashed line. Using a
combination of error cut and star/galaxy classifier ($>$0.8) given by
SExtractor (\cite{bertin96}) most background galaxies can be rejected
from the data set (dotted-dashed line). }
\label{f:hdfage}
\end{figure}

To account for the background contamination we applied different
correction procedures. For all of them we determined the age
distribution for the HDF-S objects (Fig. \ref{f:hdfage}), using the
procedure given in Sect. \ref{s:cumulative}. Due to the different
limiting K-band magnitude in the observed systems the age distribution
of the HDF was determined for each globular cluster system separately.

In the first trial (see Paper III) we corrected age distribution for
the observed systems following a worst case scenario. Hereby we
subtracted the number of HDF-S objects in each age interval (older
than {\bf 'X'}) from the number of objects found in the observed
systems. This works only in cluster samples much larger than the HDF-S
sample, but will fail in less numerous systems, e.g. NGC~3115,
NGC~7192 and NGC~4478. 

\begin{figure}[]
\centering
\includegraphics[bb=50 430 580 710,width=8cm]{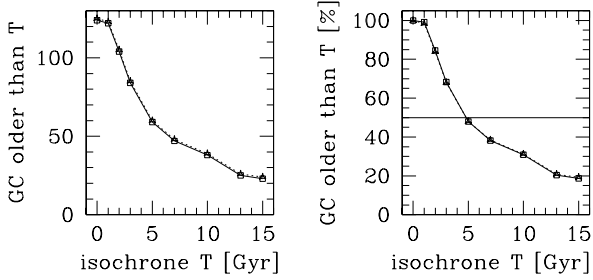}
\caption{Cumulative age distribution in NGC~5846. The left panel shows
the cumulative age distribution before (triangles, dotted line) and
after (open squares, solid line) correction for contamination applying
an error cut as well as a selection by star/galaxy classifier. In the
right panel the distribution has been normalised to the total number
of objects in the sample after correction (125 objects within the
colour limits). The age dating was done following the SSP models by
Bruzual \& Charlot.}
\label{f:cum5846}
\end{figure}

\begin{figure}[]
\centering
\includegraphics[bb=50 350 580 710,width=8cm]{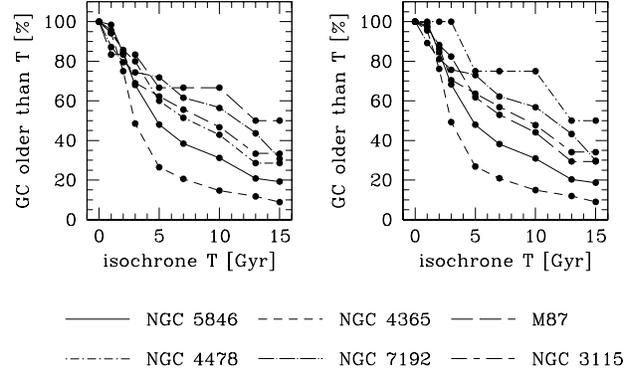}
\caption{Cumulative age distribution (using BC00) in the globular
cluster systems observed so far. The left panel shows the relative
distribution without background correction. The correction applied in
the right panel follows the procedure given in the previous section.}
\label{f:galaxies}
\end{figure}

In a second run the correction was applied to the colour-colour
distribution in the observational data. Before counting objects with
respect to a given model isochrone we rejected all objects in the
observed sample if a HDF-S object was found with a difference in
$(V-K)$ and $(V-I)$ $<$0.1 mag.  This links each contaminating object
directly to an observed object. Differences in the colour distribution
are therefore included. Observed systems with a large number of
objects are handicapped in this procedure if a background object is
found in the vicinity of several objects in the observed cluster
system. Additionally no informations about the shape of the background
objects are included yet. The globular cluster samples are selected for point
sources whether the HDF objects include as well (even mostly) extended
objects.

\begin{figure*}[]
\centering
\includegraphics[width=8cm]{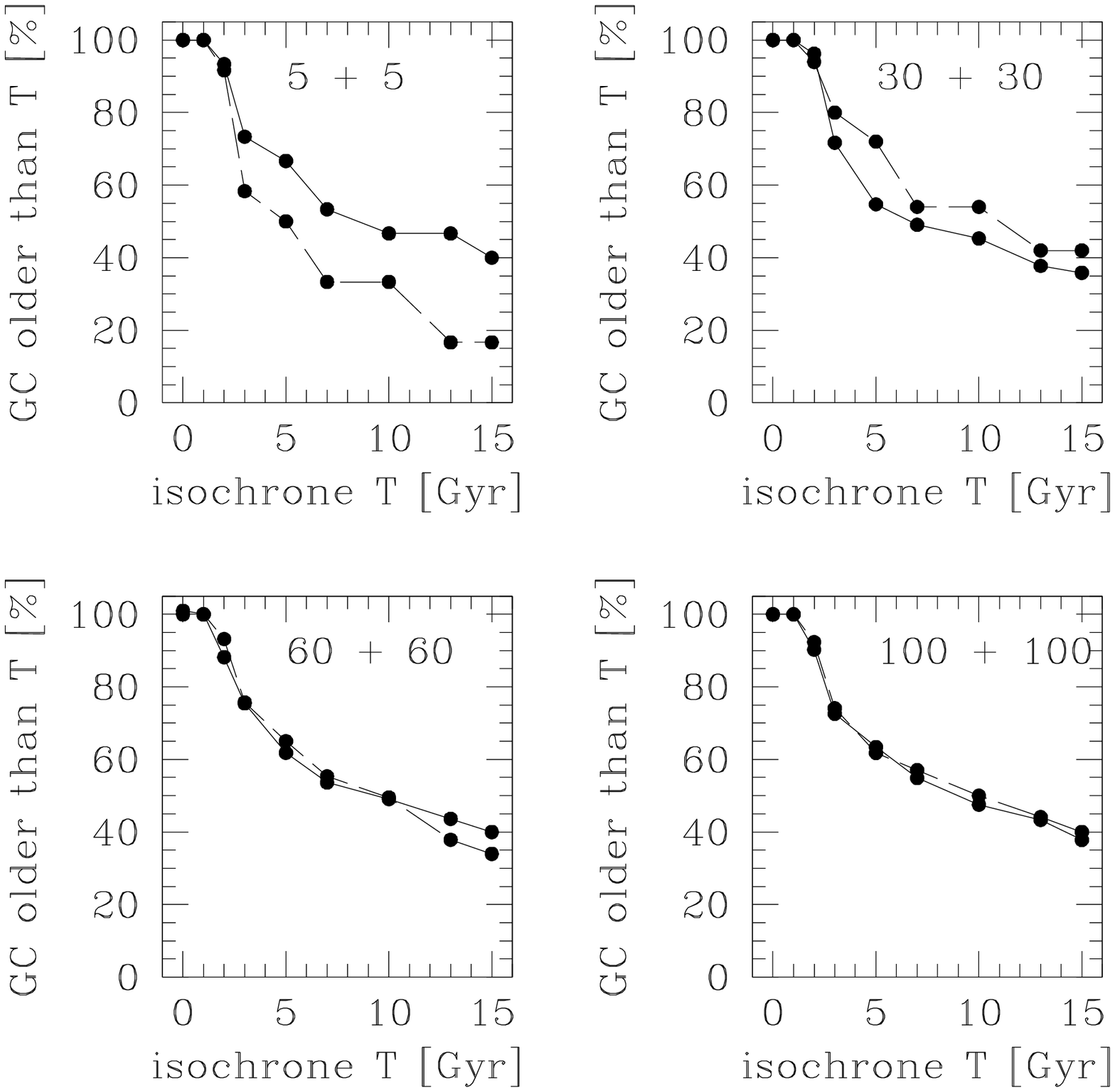}
\includegraphics[width=8cm]{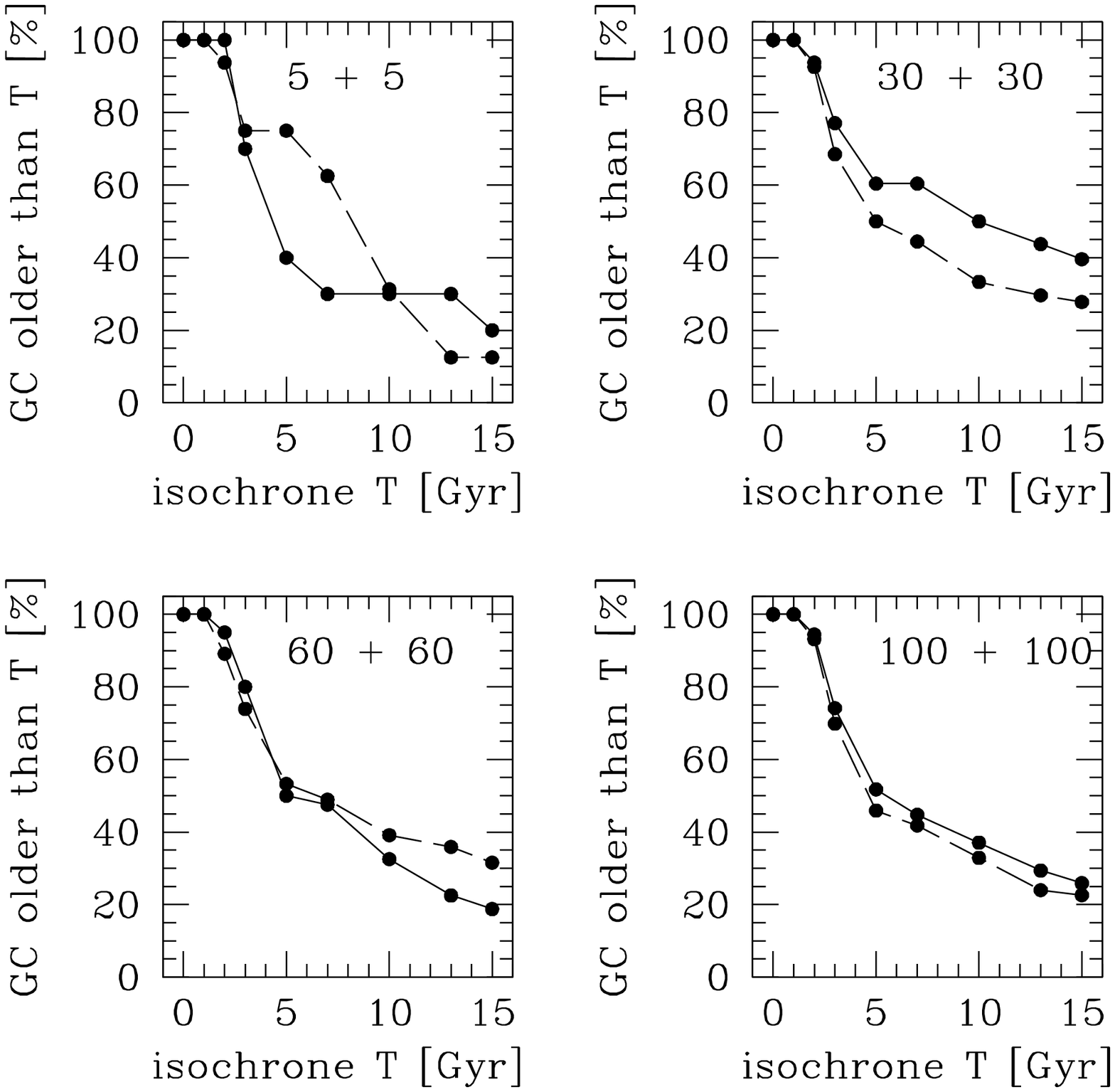}
\caption{Cumulative age distribution for modeled samples, consisting
of a different number of objects but assuming a constant mixture of
old and intermediate age objects (50$\%:50\%$). The results are shown
for the BC00 (left) and VA99 (right) SSP models respectively. Each
panel shows the age distribution for two out of six different runs of
modeling. }
\label{f:stability1}
\end{figure*}

The final most realistic correction for background objects was
therefore done by setting limits for the photometric error in all
bands (0.15~mag) and the parameter ``classifier'' given by
SExtractor. Due to the extraordinary depth of the HDF images the
classifier has been set to a limit of 0.8, which should be sufficient
to reject most of the extended objects. As an example the correction
effect in the cumulative age distribution for NGC~5846 is shown in
Fig.
\ref{f:cum5846}. In the left panel the absolute number of objects per
age bin [0 Gyr, T] before and after correction is shown. The right
panel gives the relative age distribution for both sets. In case of
NGC~5846 we find only 7 HDF-S objects within the selected colour
range, which results in an insignificant correction. Nevertheless -
this changes drastically for smaller cluster samples or in case of
deeper imaging of the globular cluster systems.

\begin{figure*}[]
\centering
\includegraphics[bb=50 480 580 710,width=12cm]{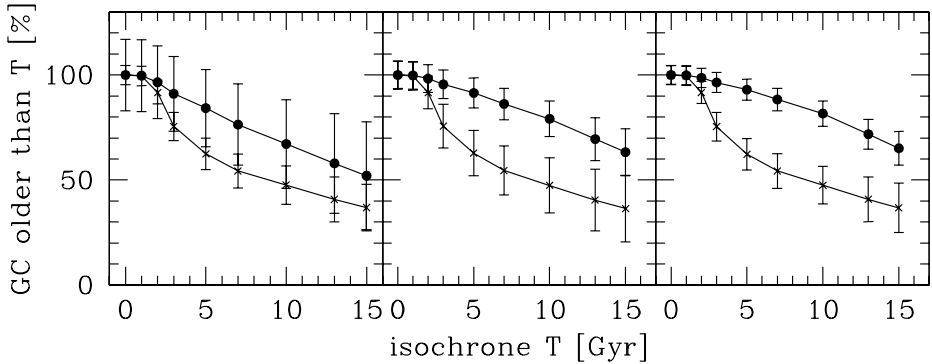}
\caption{Cumulative age distribution (using BC00) derived for models
of 10, 60 and 120 objects respectively. To show the importance of
large number of objects we show the standard deviation of the counting
rate. The models were created for a purely 15 Gyr old population
(solid circles) and a 50$\%$: 50$\%$ mix ($\times$) of 15 and 3 Gyr
old objects.}
\label{f:stats_test}
\end{figure*}

\begin{figure*}[]
\centering
\includegraphics[width=8cm]{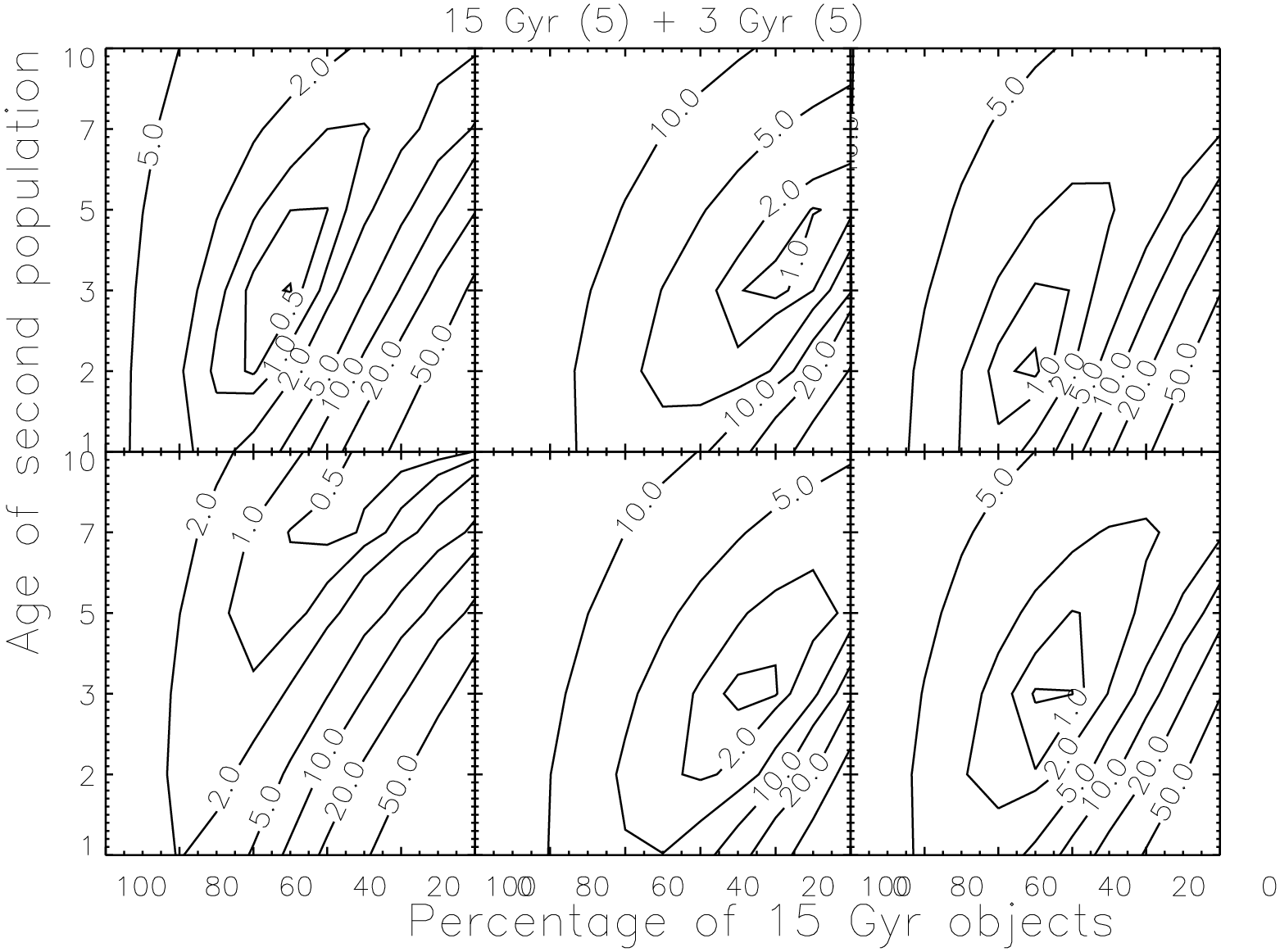}
\includegraphics[width=8cm]{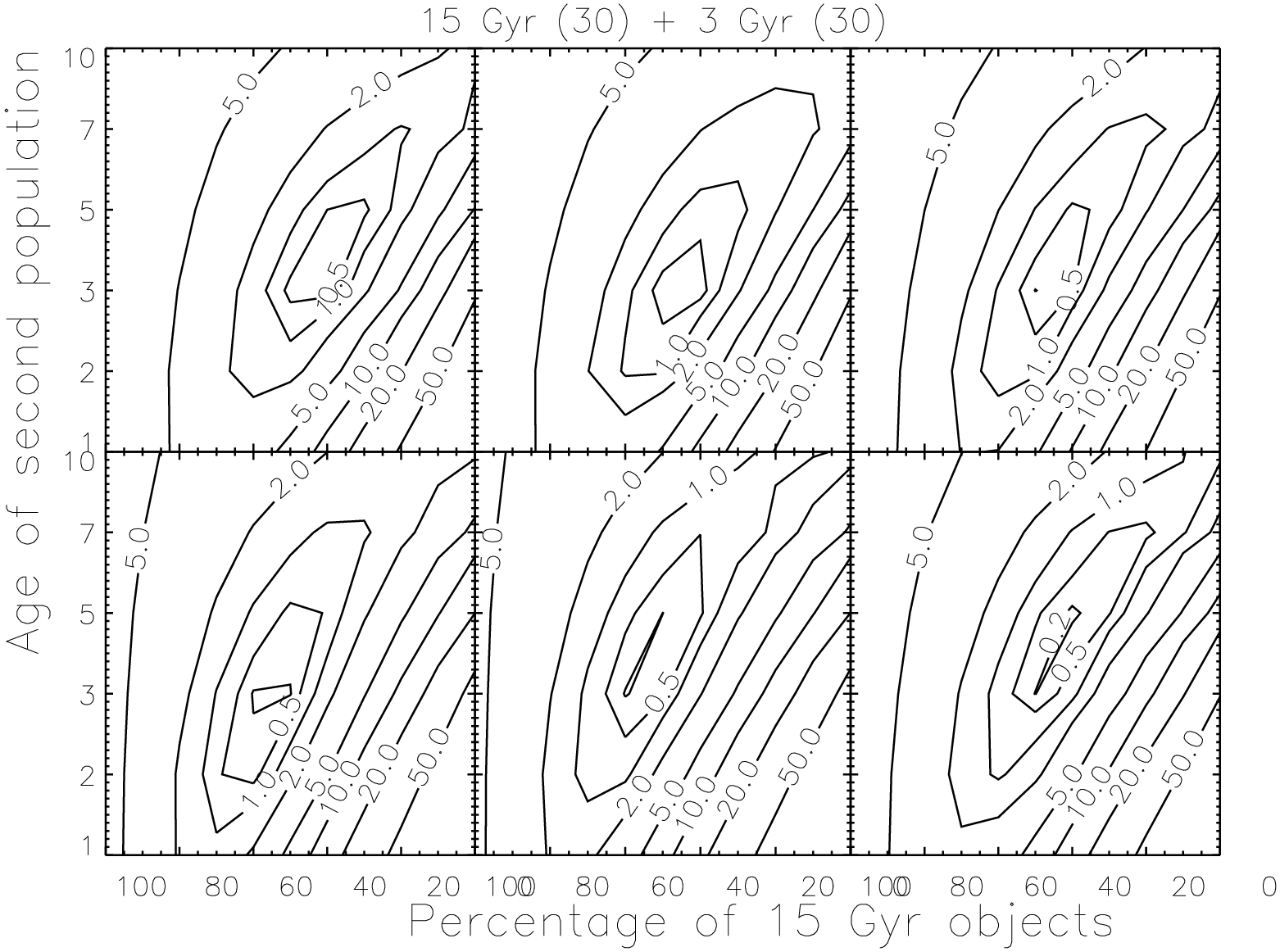}
\includegraphics[width=8cm]{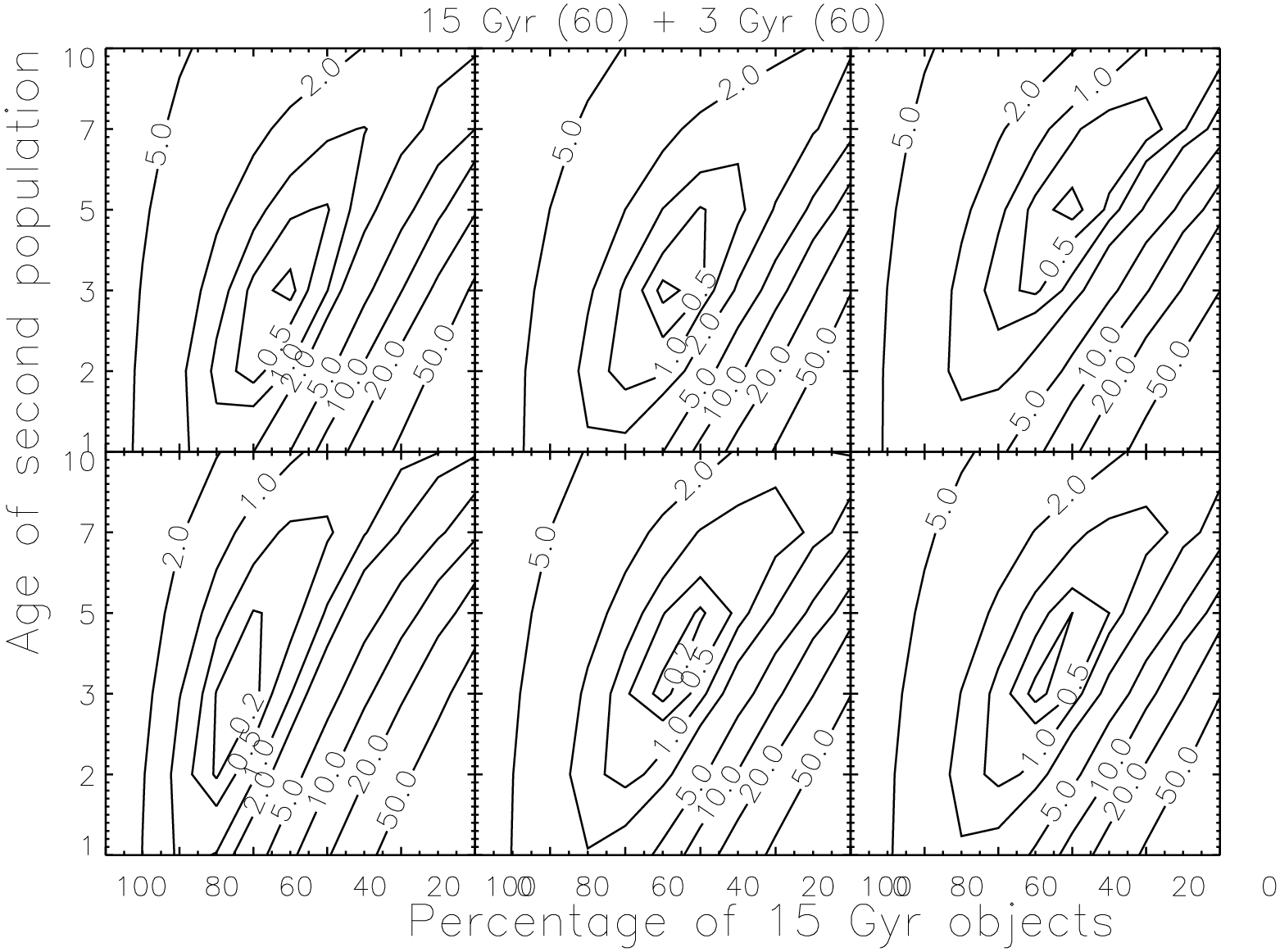}
\includegraphics[width=8cm]{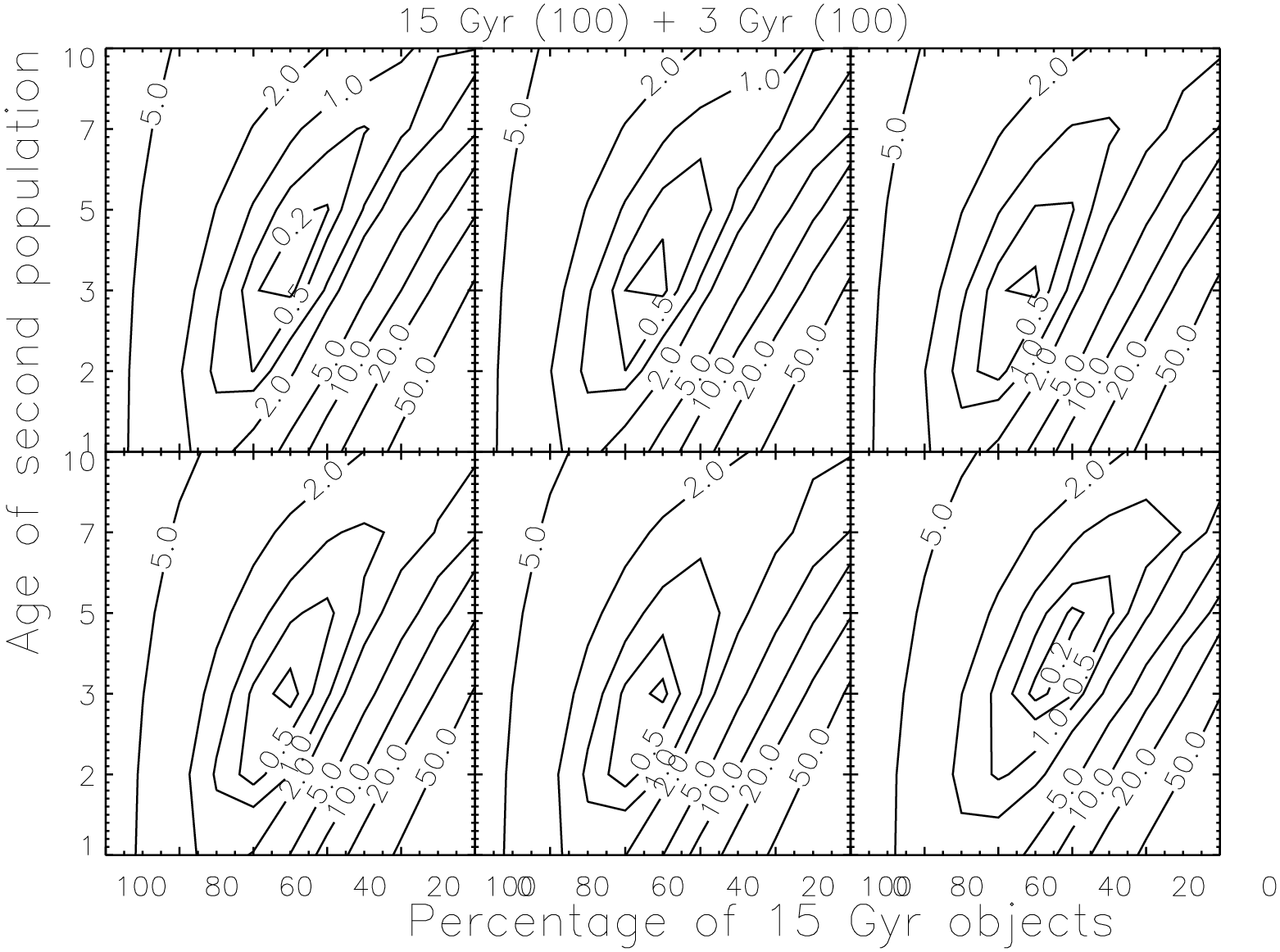}
\caption{Examples of the reduced $\chi$$^2$ test ( see section 3.2) results for
different sample sizes (10, 60, 120 and 200objects). The age
distribution for a mixture of 50$\%$ 15 Gyr and 50$\%$ 3 Gyr old
clusters (see Fig.\ref{f:stability1}) with the complete set of models
are compared. On top of each plot the number of old and intermediate
objects within the ``red'' population is given. We compare the results
for six separate runs of modeling.}
\label{f:stability2}
\end{figure*}

In our comparison of cumulative age distributions we work with
globular cluster systems of very different sizes, NGC~5846 and
NGC~4365 being the largest with a total number of about 190 clusters,
and on the other side NGC~4478 with only 21 GCs, most of them excluded
from the age distribution by the colour cuts. As described in Sect.
\ref{s:modelling} we used NGC~5846 as template system and have
therefore to check whether the age distribution and the resulting
$\chi$$^2$ test depend on the size of the observed globular cluster
systems. In Fig. \ref{f:galaxies} we show the cumulative age
distribution for all observed globular cluster systems before and
after correcting for background contamination. Due to the selection
criteria for background objects the correction effect is rather small,
except for NGC~4478, where the remaining cluster set might not allow
statistical stable results (see next section).

\begin{figure*}[]
\includegraphics[width=8cm]{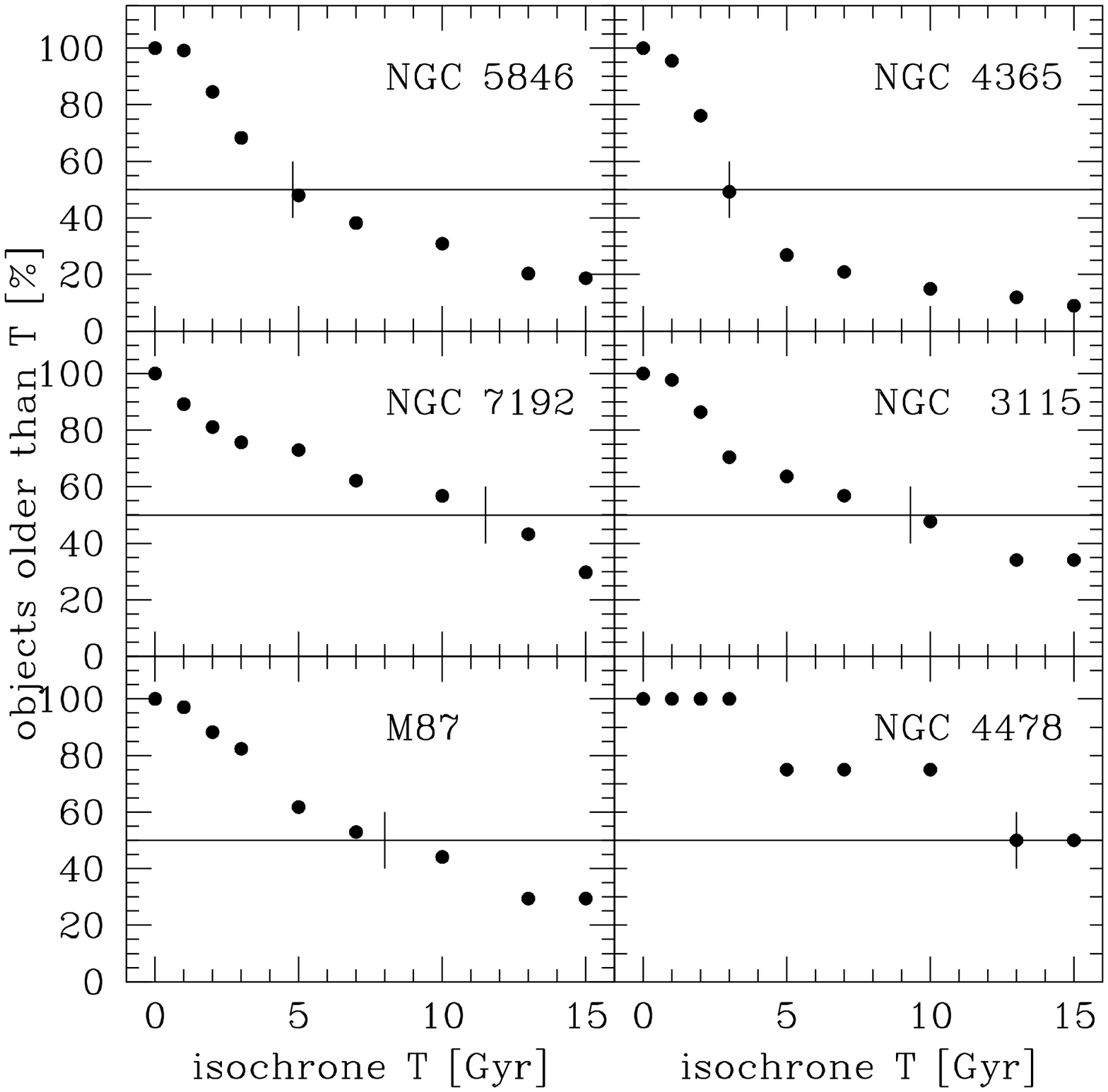}
\includegraphics[width=8cm]{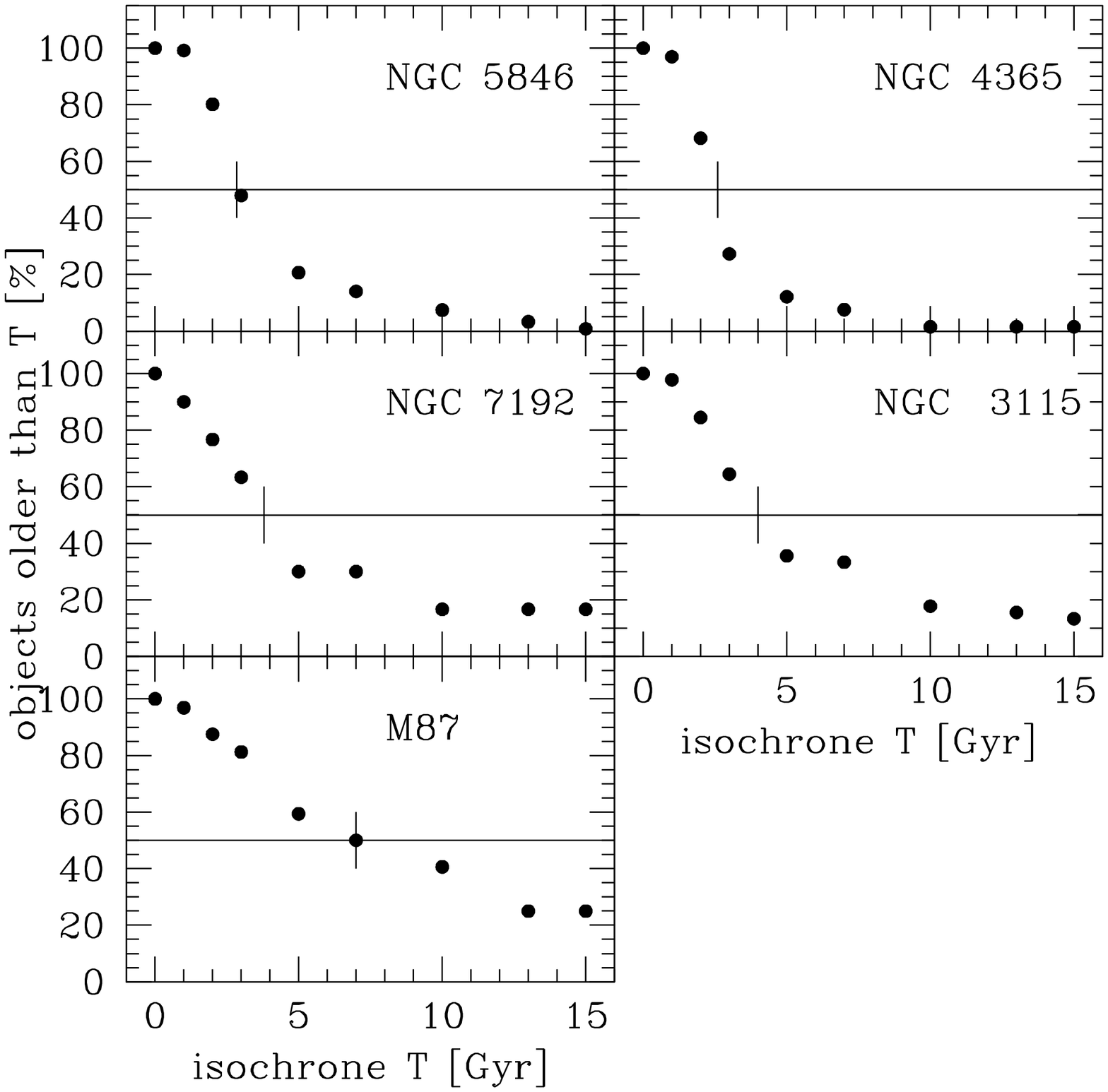}
\caption{Cumulative age distribution in the globular cluster systems
using BC00 (left) and VA99 (right) SSP model isochrones. The samples
are corrected for background contamination following the procedure
given in section \ref{s:contamination}. The horizontal line marks the
50$\%$ level and the vertical line its intersection with the derived age
distribution.}
\label{f:galaxy_dis}
\end{figure*}

\subsection{Stability of contamination correction}
\label{s:stability}

In the previous section we opened the discussion to which extend the
sample size affects the final result. In the competition between
spectroscopy and photometry toward relative age dating this becomes
even more important, the sample size being the advantage of
photometry. But what is the minimum size of the cluster system, for
which the scattering in the data (e.g. due to photometric errors)
becomes negligible in the age distribution? To answer this question we
modeled the colour-colour diagram of a mixed cluster population
(50$\%$ 15 Gyr old objects and 50$\%$ 3 Gyr old objects. The
age distributions for systems between 10 and 200 objects are shown in Fig.
\ref{f:stability1}.We find that in systems with less than 60 objects 
the age distribution varies significantly. With respect to section 3.3
we find this effect being even stronger if the VA99 model isochrones are used.

In order to minimise the scattering effect in the simulated cluster
system we generated 1000 models and work with the statistical mean of
the age distribution (see Sect. \ref{s:monte}). Still the
sample size for the models has to be considered. In Fig.
\ref{f:stats_test} we compare the age distribution for models
consisting of either 100$\%$ old objects or a 50$\%$: 50$\%$ mix of 15
Gyr and 3 Gyr old globular clusters in a set of only 10, 60 or 120
clusters. The error bars give the standard deviation of the object
counts in the 1000 models. In systems with only 10 objects the age
distribution for a mixed population and a purely old system are,
within the error limits, indistinguishable. In slightly larger samples
(e.g.~60 objects) the age distributions of a purely old and a mixed
population overlap in the older age bins. The result would be multiple
findings of best fitting models (see Sect. \ref{s:isthere}). As a
last point we want to mention that our primary assumption of a random
distribution in $(V-K)$ becomes critical for small samples.

\subsection{Is there a significant intermediate age cluster population?}
\label{s:isthere}

In Figure \ref{f:stability2} we show the result of the reduced
$\chi$$^2$ test for finding a specific input model within our complete
set of 66 models of varying sample sizes for the BC00 model. In
contrast to the simulations the age distribution in the observed
globular cluster systems represents a single sample still subject to
stochastic effects. In order to test the stability of our derived
ages/number ratios for various sample sizes we compare the age
distribution for six individual simulations (assuming the same model
parameters but different sample sizes). The example shown here are
based on the BC00 model. As we can see, we should be able to detect
sub-populations, if their ages differ by several Gyr. Nevertheless,
for small sample sets the results become quite unstable, i.e. the age
of the young sub-population and the size ratio between both
populations are more uncertain. Again- this is more important for the
VA99 model, due to the steeper $(V-I)$ gradient of the isochrones. We
also have to keep in mind that all the models were created based on
our largest globular cluster system- NGC~5846 and we therefore expect
the most reliable results for sample sets which are in size and colour
range similar to NGC~5846. Special attention has to be paid to
NGC~4478, where the majority of objects is found in the blue colour
range. In order to draw stricter conclusions about the red
sub-populations the modeling parameters (size, $(V-K)$ colour range)
have to be chosen with respect to the observed cluster system.

Although we have discussed the effect that unresolved background
objects can have on the results, we are aware that we are at this
point not able to offer a quantitative correction for each of the
observed systems. Even if our final selection criteria (error cut and
classifier) for background objects (see Sect. \ref{s:contamination})
are too strict - we do not expect contamination to play a major role
in large globular cluster systems. Only the NGC~4478 (21 cluster in
total) sample falls into the low number region where the stability of
the $\chi$$^2$ test is questionable.

\begin{figure*}[]
\centering
\includegraphics[width=8cm]{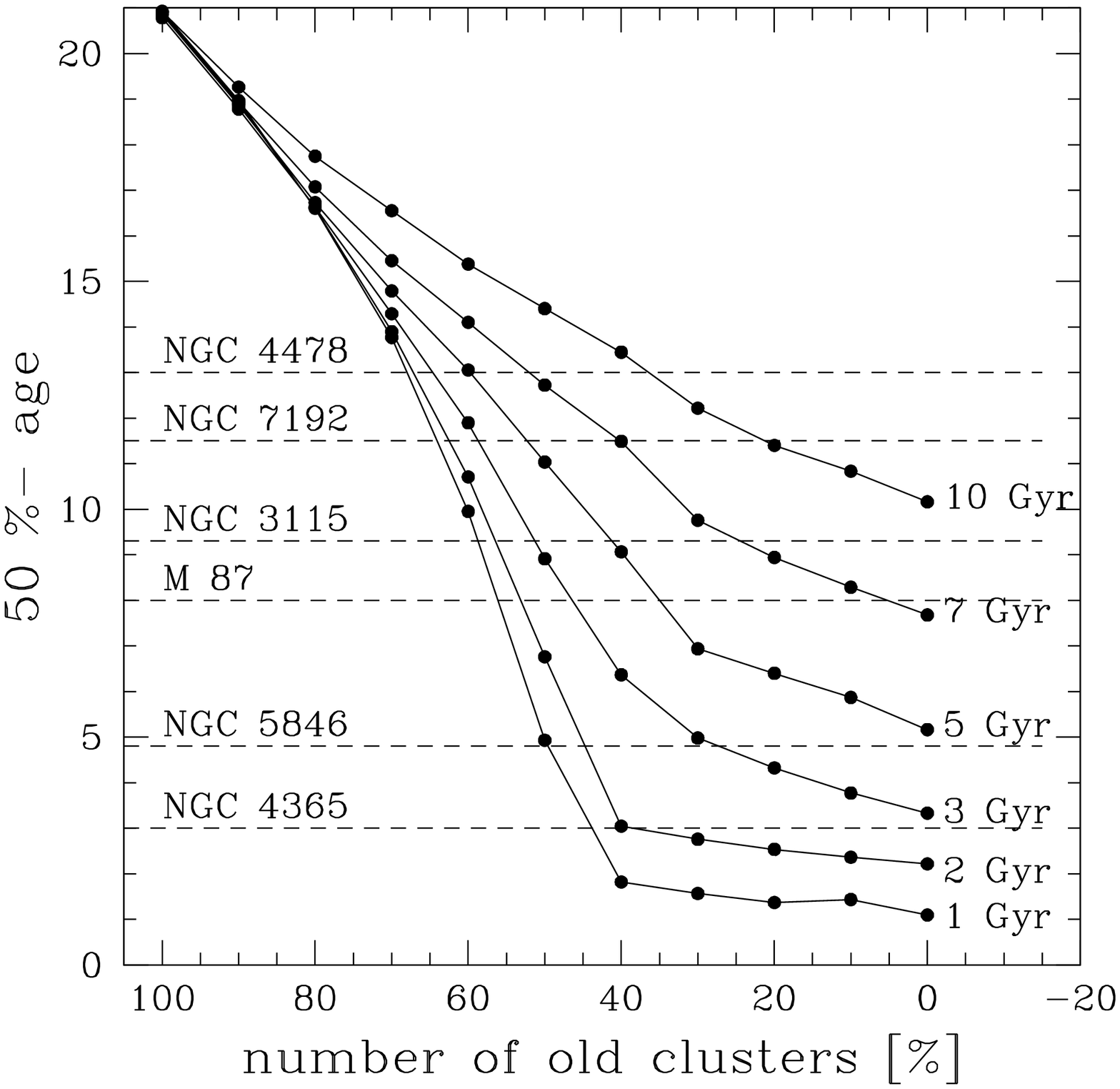}
\includegraphics[width=8cm]{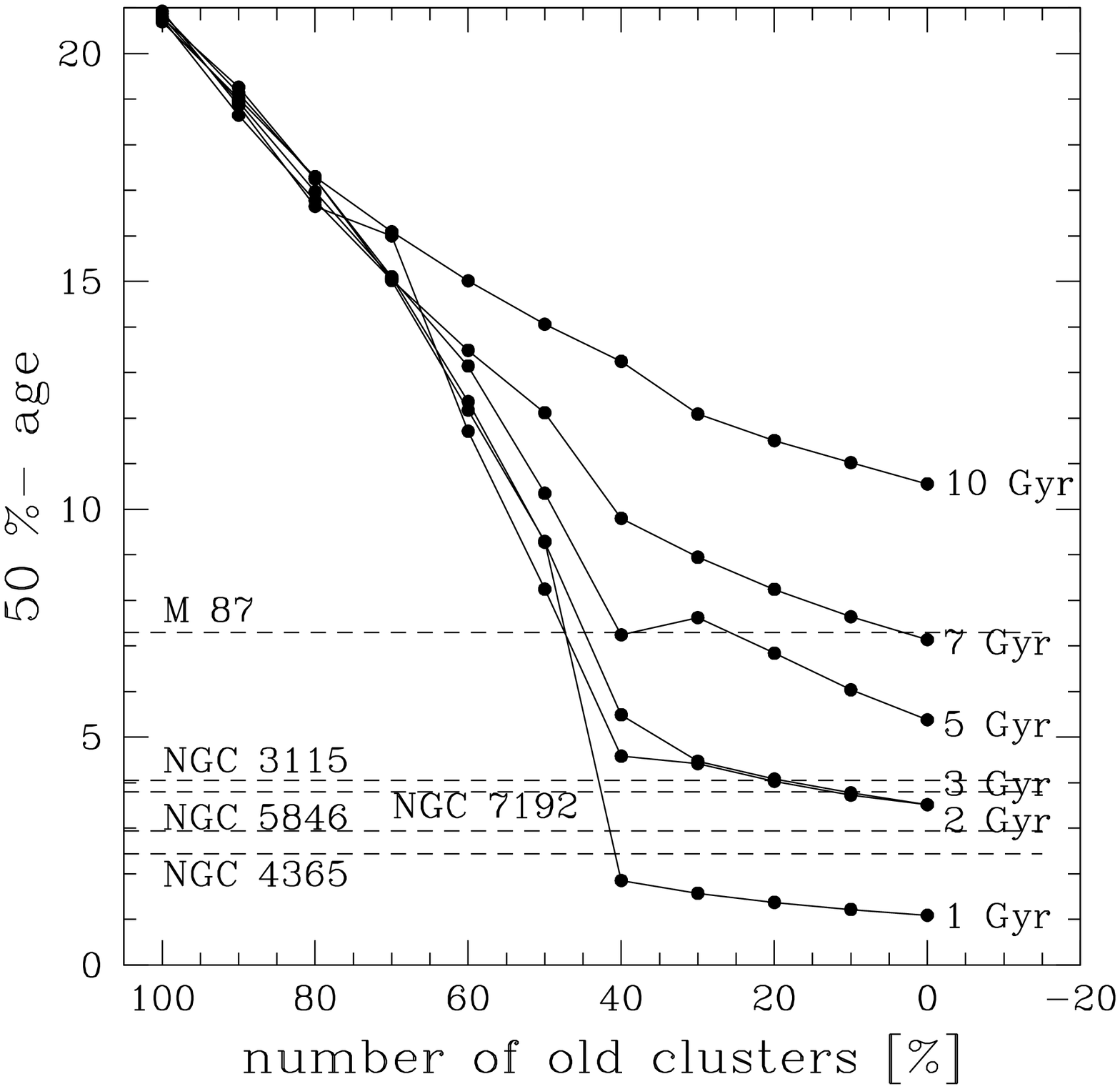}
\caption{We  compare the 50$\%$-age calculated by a best fit of the
cumulative age distribution for observed and simulated systems using
BC00 (left ) and VA99 (right) SSP model isochrones. The
x-axis gives the amount of 15 Gyr old objects within the model and the
label on the right hand side the age of the intermediate age
sub-population. The larger spread in $(V-I)$ for VA99 isochrones
results in a less reliable result for smaller samples. All systems
have been corrected for background contamination.}
\label{f:50percent}
\end{figure*}

\begin{figure*}[]
\includegraphics[bb=50 430 650 710,width=19cm]{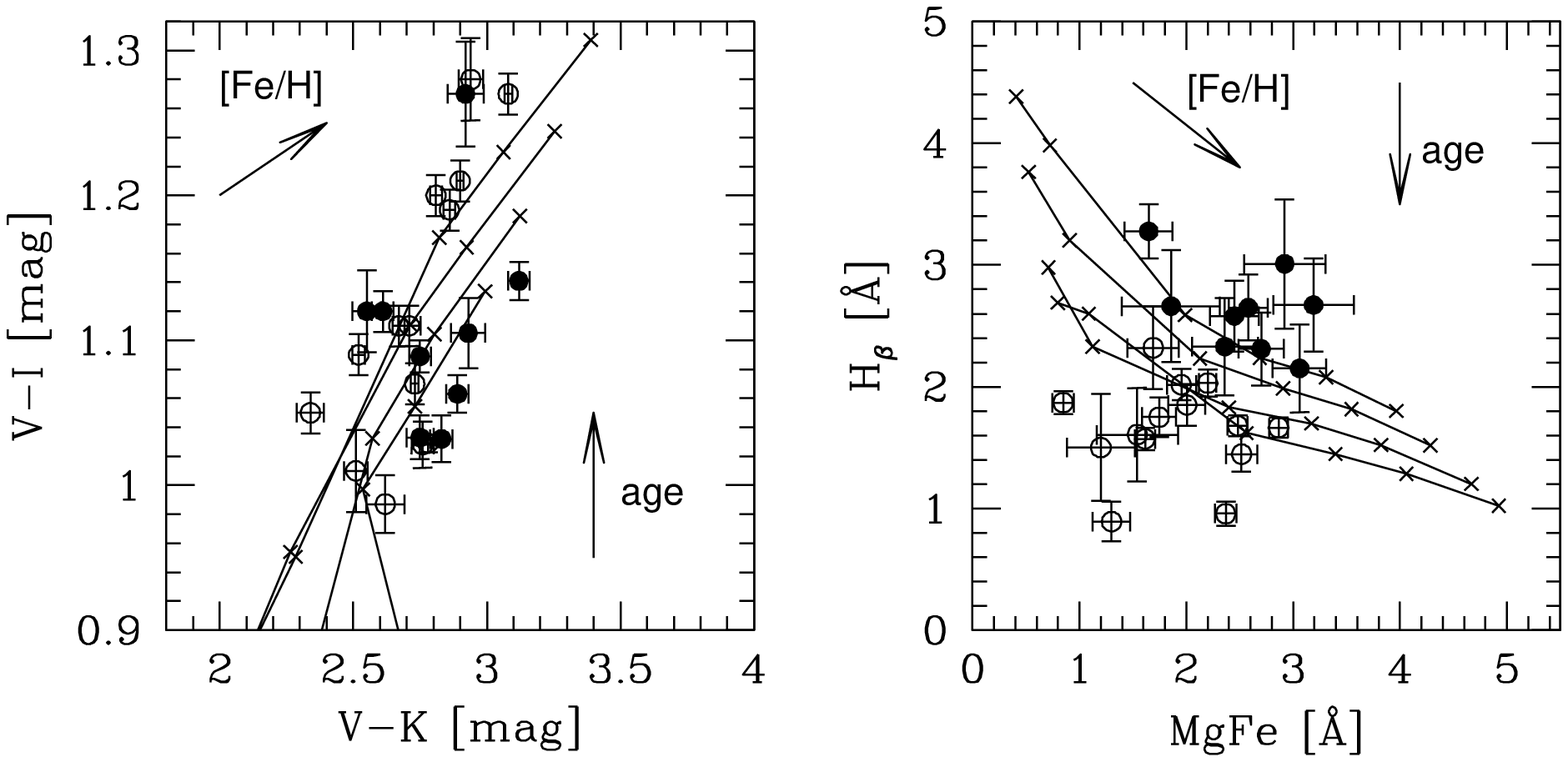}
\caption{Left panel: Colour-colour diagram for globular clusters
 in 4 different galaxies. The model grid follows the Maraston (2001)
 SSP model isochrones (ages between 3 and 15 Gyr following the age-
 arrow). The metallicity ([Fe/H] between -2.25 and +0.35) increases in
 direction of the metallicity arrow. Right panel: Comparison of index
 measurements with the models by Thomas et al.2003 (age range: 3- ~15
 Gyr, metallicity range: -2.25-~+0.67). Objects with spectroscopic
 ages below 5 Gyr are shown by solid circles.}
\label{f:phot_spec}
\end{figure*}

\section{Results for observed systems}
\label{s:results}

The goal of relative age dating via the cumulative age distribution is
to find evidence for various major star formation events in elliptical
galaxies and to constrain those in their age difference and importance
for the galaxy evolution. To do so we compare two features of the age
distribution, the so-called 50$\%$-age and the result of an
$\chi$$^2$-test between simulated and observed globular cluster
systems. The 50$\%$-age is defined as the age at which the
cumulative age distribution intersects with the 50$\%$ level. Figure
\ref{f:galaxy_dis} shows the cumulative age distribution for all globular cluster systems
observed so far, using the SSP model by Bruzual \& Charlot and
Vazdekis, respectively.

If we compare this 50$\%$-level age derived for the observed systems
with the models (see Fig. \ref{f:50percent}) we find a significant
model dependence. We will include this parameter in our
discussion.

For NGC~5846 and NGC~4365 with large globular cluster systems we
obtain consistent results, independently on the SSP model. The
50$\%$-age of both systems agrees with the 50$\%$-age derived for a
mixed population, hosting a second cluster population at an age up to
5 Gyr. In M87 the 50$\%$- age gives as well evidence for a second
population with an age of $\sim$7 Gyr. For NGC~7192, NGC~3115,
NGC~4478 the results for the 50$\%$-age are inconsistent and depend on
the used SSP models. NGC~4478 is a exceptional case for this
study. The observed globular cluster system is very small and consists
of mostly metal-poor objects. Within the $(V-K)$ colour range we find
only 3 clusters, which is not sufficient for our analysis.

As mentioned in Paper III our intention regarding the modeling was the
possibility to make quantitative statements about the star formation
history in various galaxies. In the next section we present the result
of the comparison between of the observed globular cluster systems and
the Monte-Carlo simulations. We provide some limits on the age
structure for 4 of the observed systems. As introduced in section 4.3 we
discuss the results in the light of their SSP model dependence, which
seems to be predictable in a way that applying VA99 models results in
slightly younger ages and requires a larger sample size.

\subsection{NGC~5846 and NGC~4365}
\label{s:n5846_n4365}

Spectroscopic (\cite{larsen03}) and photometric (Paper II and III)
results for NGC~5846 and NGC~4365 have shown that both globular
cluster systems host an intermediate age population indicating a
second star formation period. In Fig.\ref{f:phot_spec} we present the
photometric and spectroscopic data for the globular clusters in
NGC~4365, NGC~5846, NGC~7192 and NGC~3115 for which optical/
near-infrared photometric and spectroscopic data are available (33
objects). The left panel shows the $(V-I)$$vs.$$(V-K)$ colour-colour
distribution of all clusters compared to the MA01 model isochrones. In
the right panel the index measurements for H$_{\beta}$ and MgFe are
compared with the SSP model by Thomas et al.  (\cite{thomas03}) and
confirm the photometric age dating. The globular clusters assigned to
an intermediate age are those found in NGC~4365 and NGC~5846.

\begin{figure*}[]
\centering
\includegraphics[width=8cm]{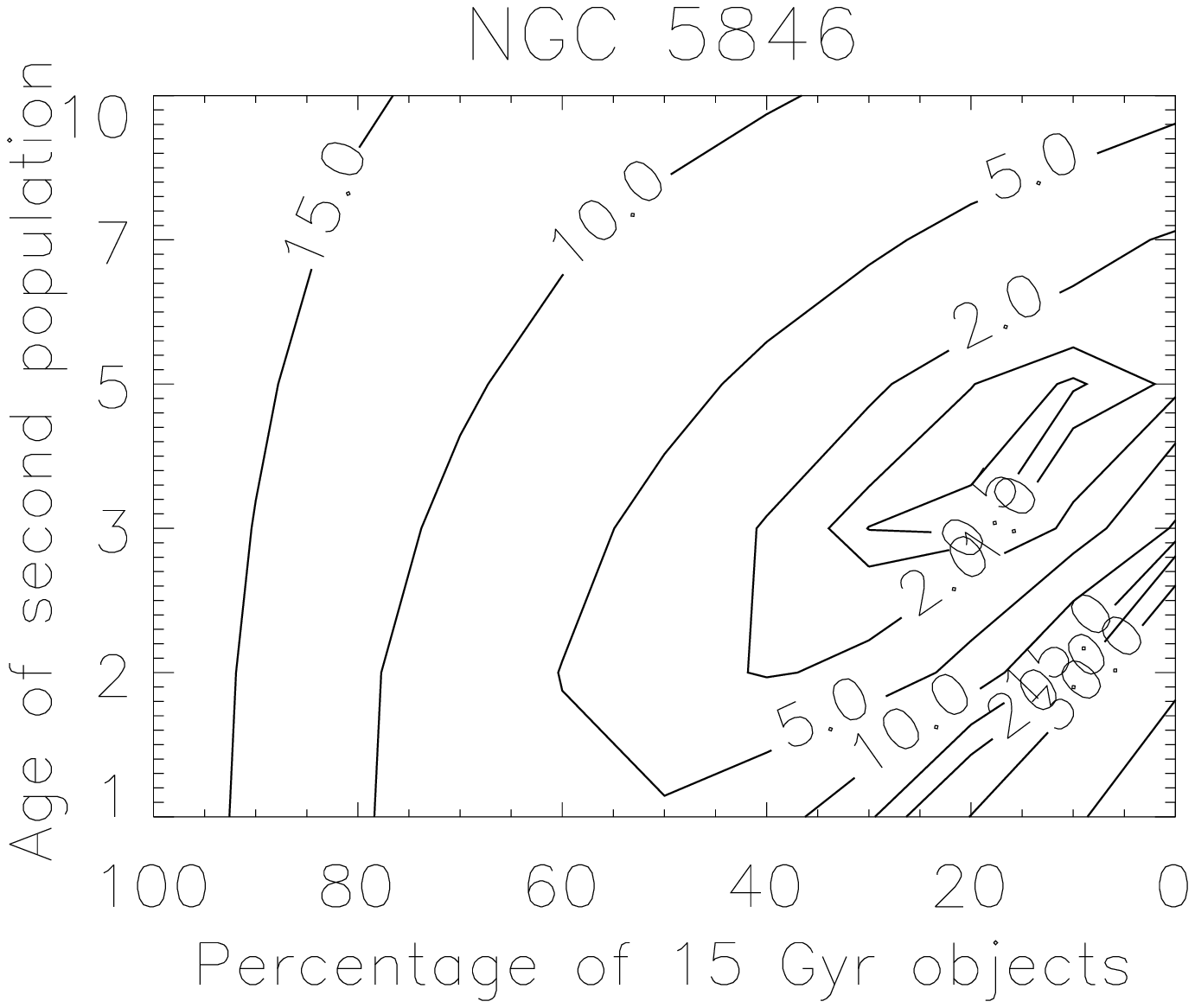}
\includegraphics[width=8cm]{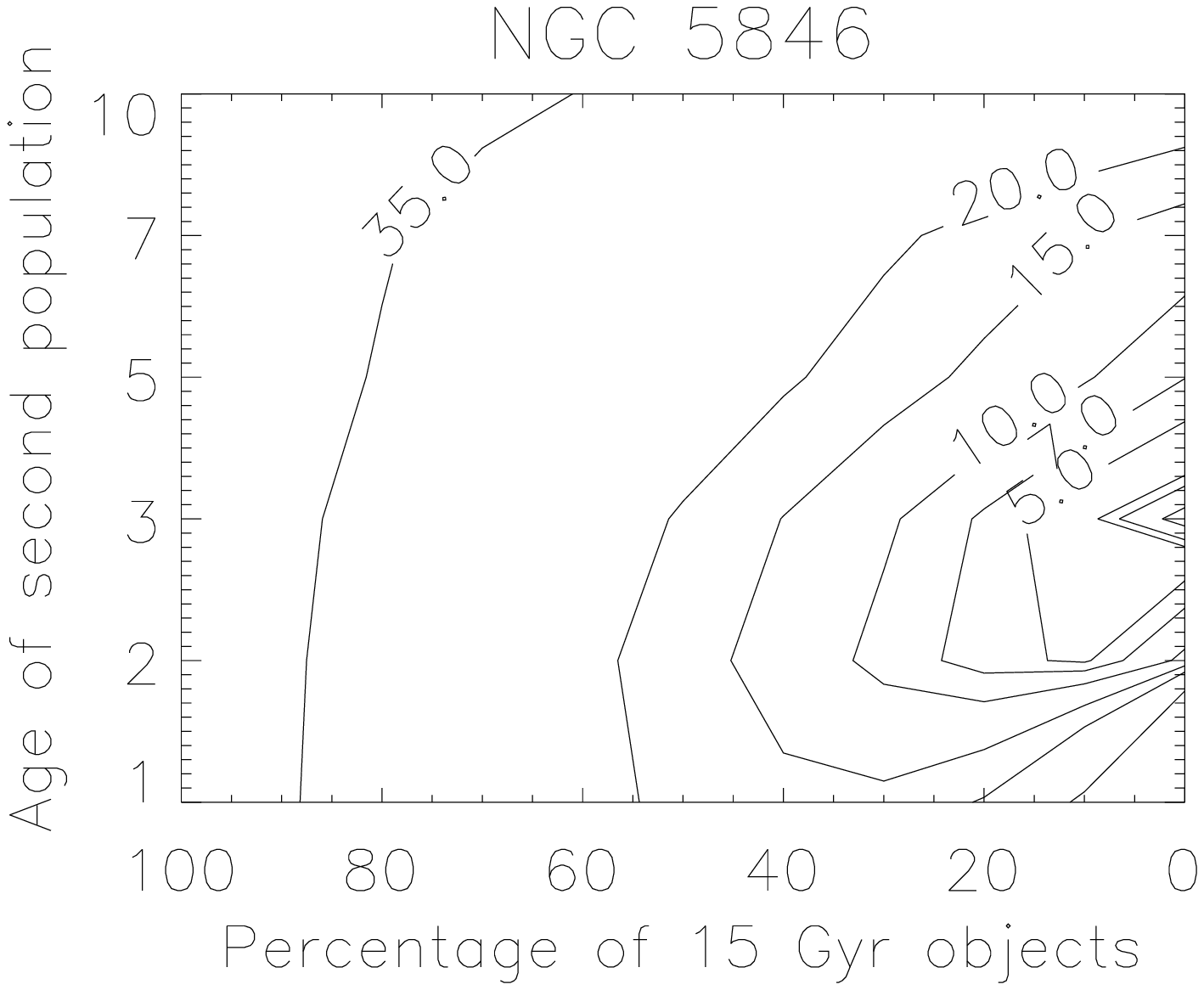}
\includegraphics[width=8cm]{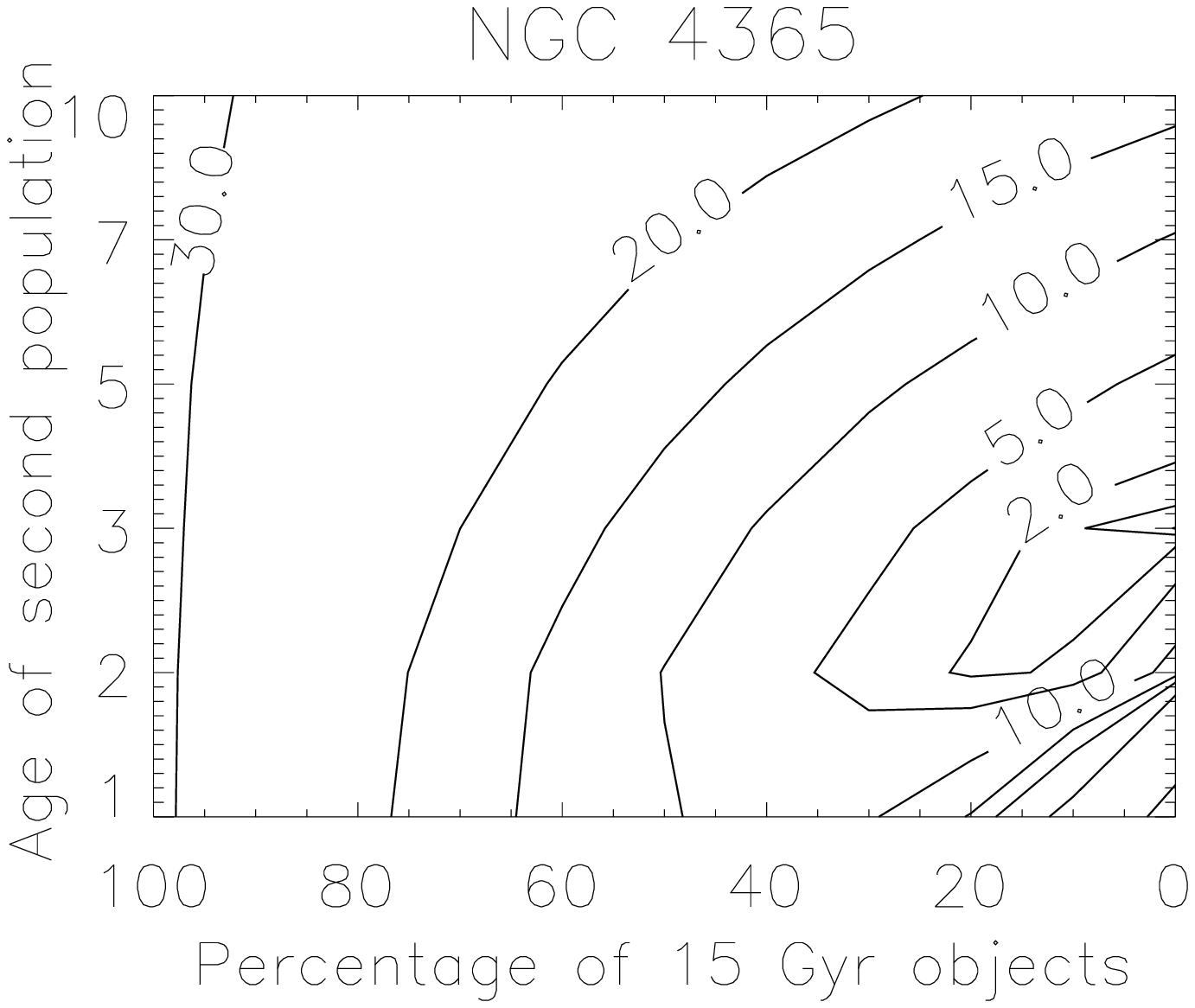}
\includegraphics[width=8cm]{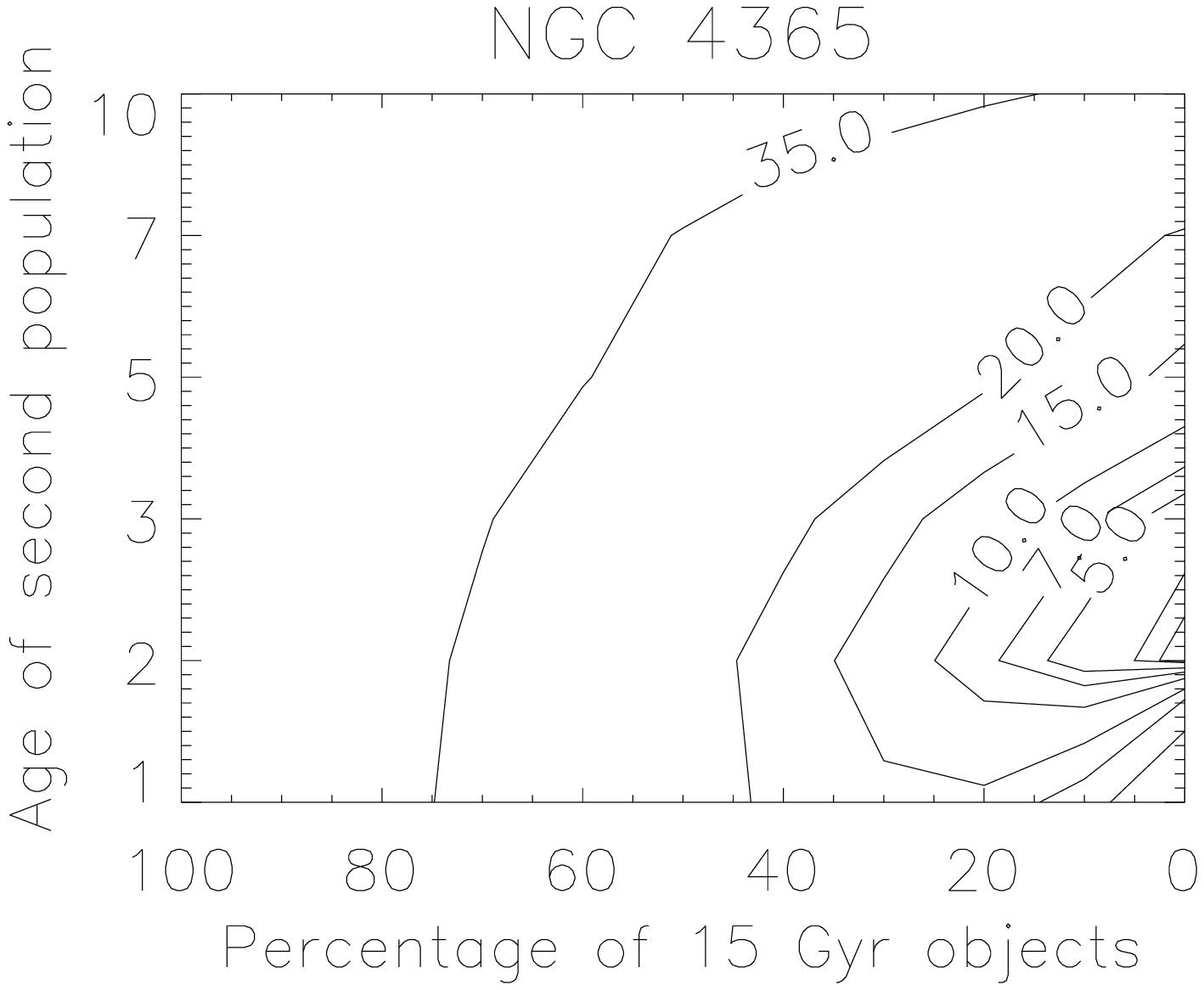}
\caption{$\chi$$^2$ test result for NGC~5846 (upper) and NGC~4365
(lower). The different levels represent the reduced $\chi$$^2$ of the
comparison between the age distribution in the observed systems and
the various models. The correction for contamination following the 
procedure given in section \ref{s:contamination} has been applied.The
left panels give the result following the BC00 models and the right
panels following the VA99 models.}
\label{f:contour1}
\end{figure*}

Consequently we expect the 50$\%$-age to show the best agreement with
the value calculated for mixed populations as well as the best fitting
model being a mixture 'old' and 'intermediate' age
globular cluster. For both SSP models we obtain 50$\%$-ages belowe 5
Gyr the Vazdekis model giving slightly younger ages. Still the results
are in good agreement with the value derived for models mixing an old
population with globular clusters with ages up to 3 Gyr.  The upper
panels of Fig. \ref{f:contour1} show the results of the $\chi$$^2$
test for NGC~5846 and gives the highest agreement between the cluster
sample and the models assuming a mixture of and old (15 Gyr)
population and roughly up to 90$\%$ 3-5 Gyr old clusters. We will
discuss this extremely large value, which we find for most of our
target systems in section \ref{s:conclude}. For NGC~4365 (Fig.
\ref{f:contour1}, lower panels) we derive a mixture of 15 Gyr and 1-3
Gyr old objects with a higher content of intermediate age objects. The
age range is in agreement with the spectroscopic results given by
Larsen et al.(2003). Both observed cluster samples are relatively rich
($N_{NGC~5846}$ and $N_{NGC~4365}$ are in the range of 190) and the
correction effect is relatively weak and, taking the stability tests
into account as well (see section\ref{s:stability}), stable.

\begin{figure}[]
\includegraphics[bb=50 380 650 710,width=8cm]{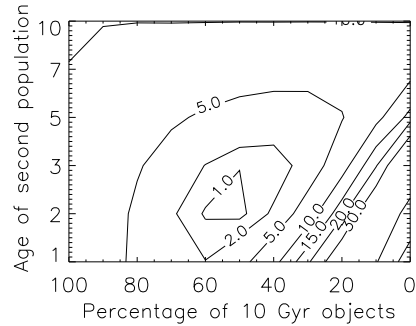}
\caption{Result of the $\chi$$^2$ test comparing the age distribution
in NGC~5846 with simulated systems assuming the {\it{old}} population
being 10 Gyr old. As clearly seen the best fitting model still
contains a much smaller, but still significant number of intermediate
age objects.}
\label{f:shift0}
\end{figure}

As mentioned in Sect. 2.1 the simulations are based on the assumption
of the old population which is 15 Gyr old. We present one example to
illustrate how the results will change in case of a 10 Gyr {\it{old}}
population (see Fig. \ref{f:shift0}). Although the relative size of
both populations change considerably, as we would expect, the
detection of a intermediate age population is confirmed and the
relative ages of both populations does not change
significantly. Moving from 15 to 10 Gyr, besides being driven by real
physical arguments, also illustrates how our results would change if
the SSP models are not perfectly calibrated.
 
\begin{figure*}[]
\centering
\includegraphics[width=8cm]{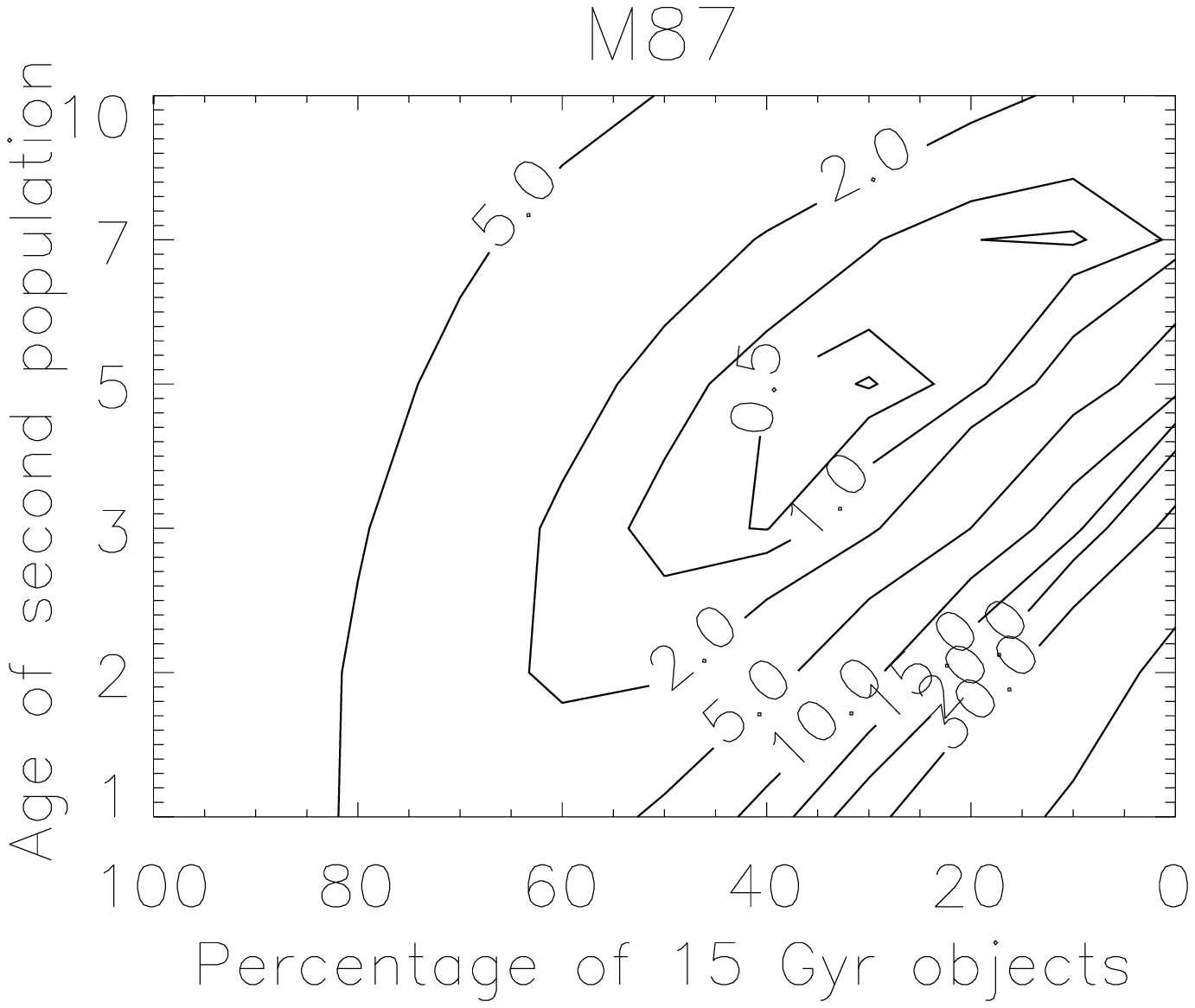}
\includegraphics[width=8cm]{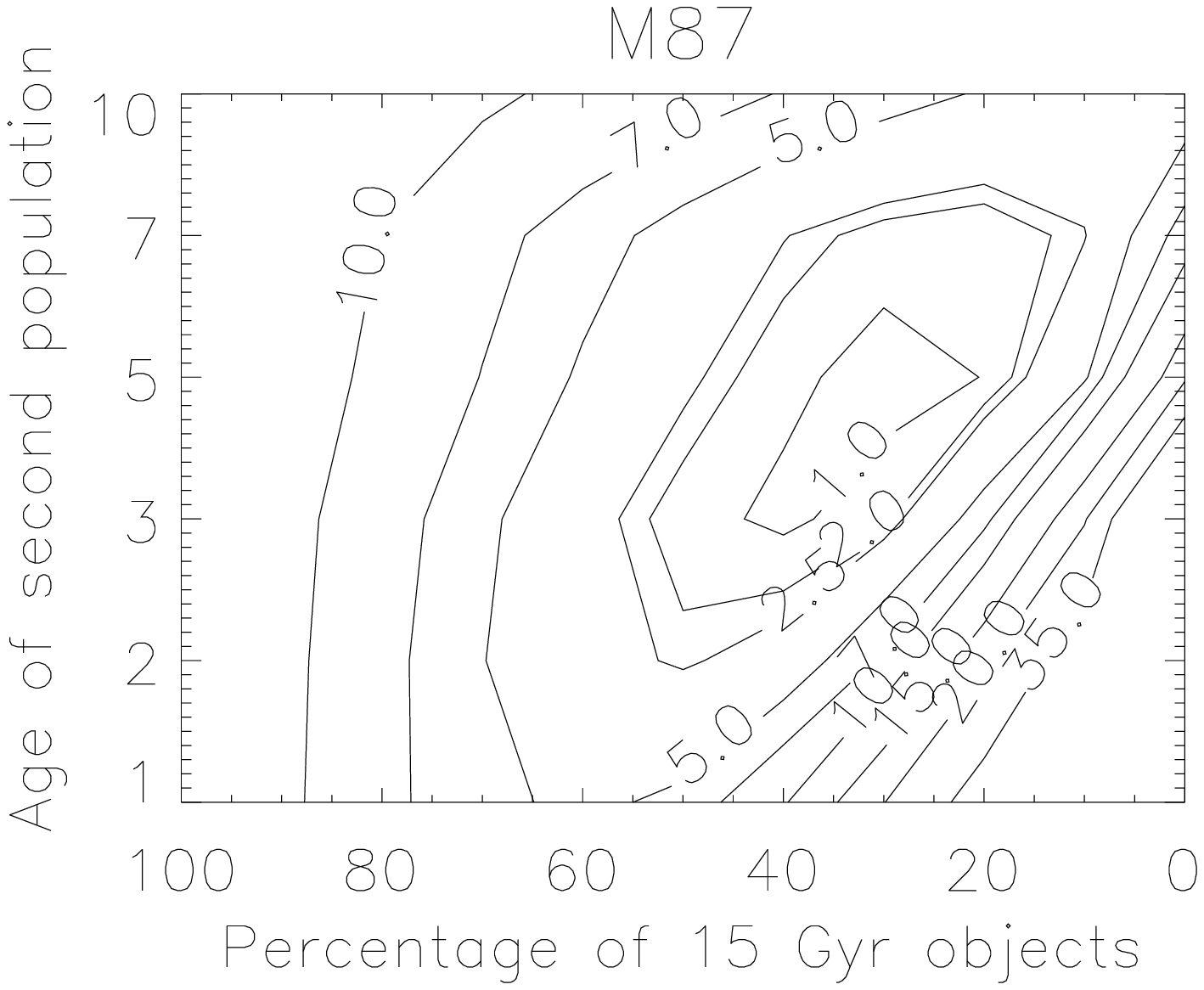}
\includegraphics[width=8cm]{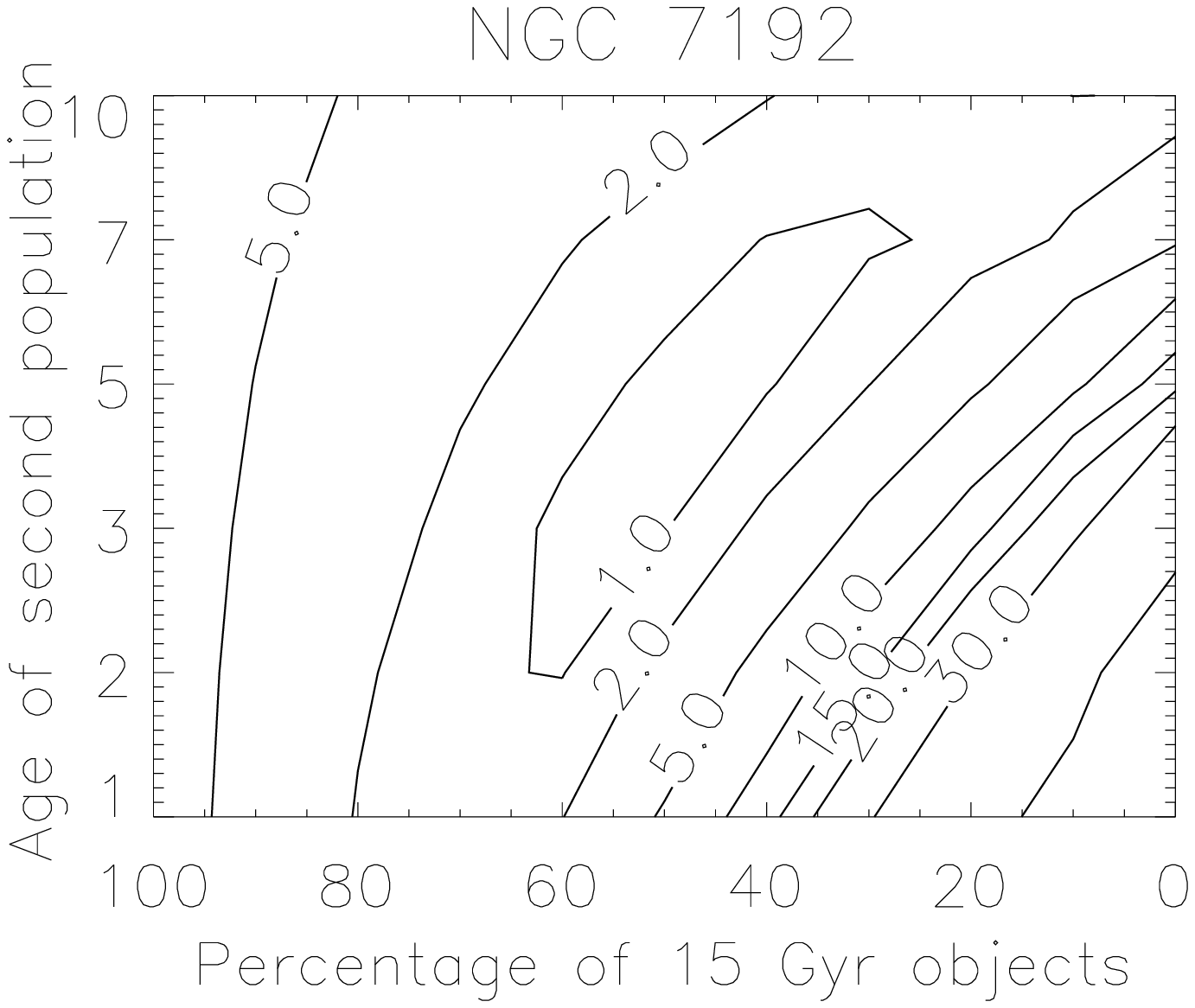}
\includegraphics[width=8cm]{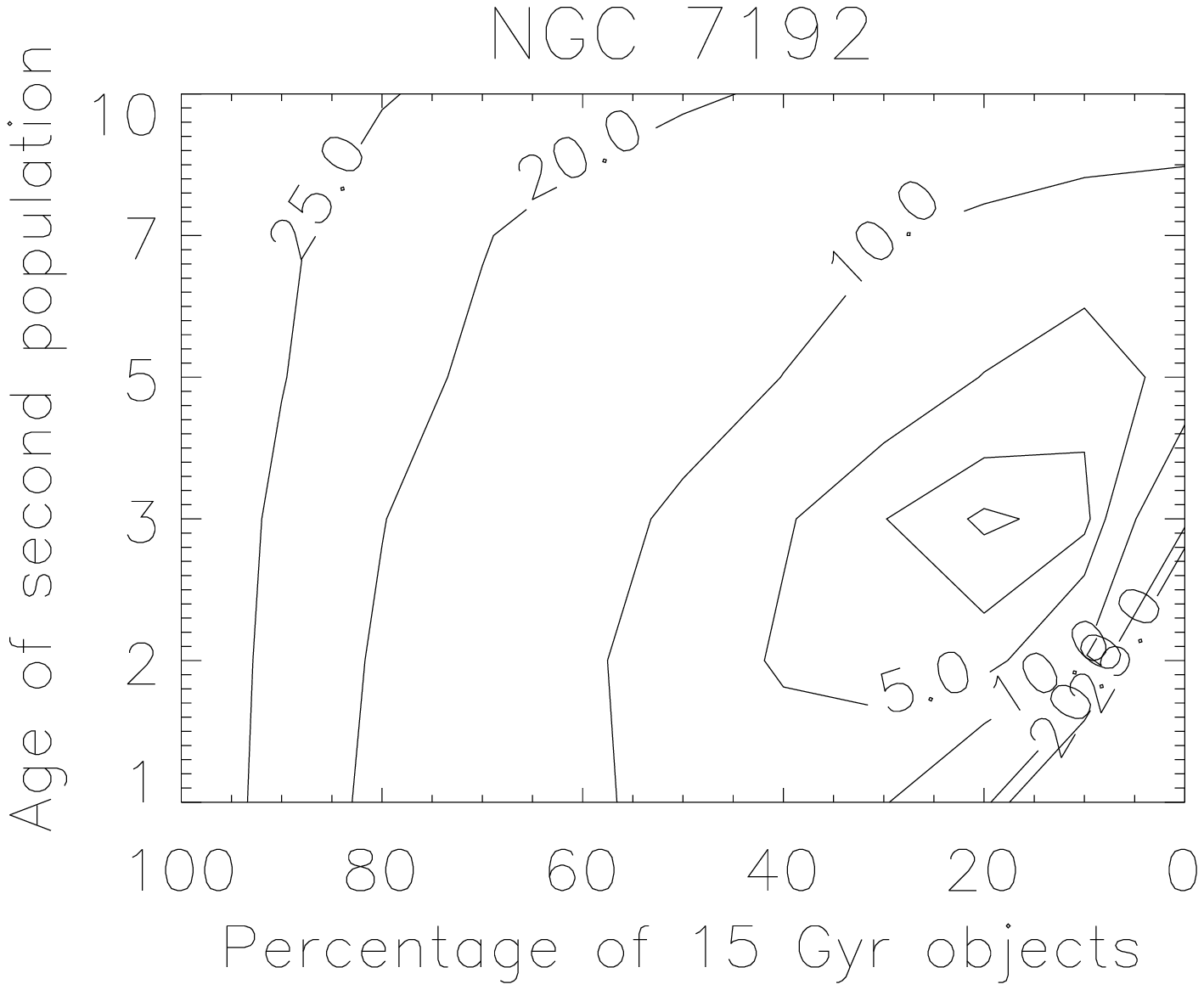}
\caption{$\chi$$^2$ test result for M87 (upper ) and NGC~7192
(lower). As in Fig. \ref{f:contour1} the results for BC00 (left
panels) and VA99 (right panels) are shown.}
\label{f:contour2}
\end{figure*}

\subsection{M87 and NGC~7192}
\label{s:m87_n7192}
With respect to the formation scenario for early-type galaxies we are
interested in the environmental effect. Therefore our galaxy sample
includes also cD galaxies (M87) and rather isolated galaxies
(NGC~7192). In both types of globular cluster systems one wouldn't
expect a second, younger globular cluster population (for M87 see
\cite{jordan02}). This expectation is only met by the
50$\%$-age derived for M87. Using the Bruzual \& Charlot model we
estimate a 50$\%$- age of 8 Gyr and for VA99 $\sim$7 Gyr, both values
being more consistent with models containing only a small population
of somewhat younger objects. The fact that this result could not be
confirmed by the $\chi$$^2$-test (see Fig. \ref{f:contour2})
led us back to the problem of small sample sizes. As soon as we
include the size of both globular cluster systems, 35 found for M87
and 39 for NGC~7192 respectively, this inconsistency can be
explained. Both systems are still not numerous enough to allow a
meaningful analysis.

\section{Discussion and conclusions}
\label{s:conclude}
\begin{figure}[h]
\includegraphics[bb=50 450 650 710,width=10cm]{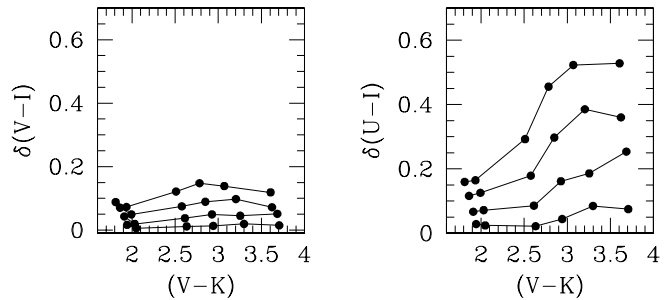}
\caption{Age resolution for SSP model isochrones by Bruzual \& 
Charlot (\cite{bruzual00}) in $(V-I)$$vs.$$(V-K)$ (left panel) and
$(U-I)$$vs.$$(V-K)$ (right panel) color-color diagrams. The age
splitting in $(U-I)$$vs.$$(V-K)$ plots is increased by a factor of 2
and will therefore result in a better age resolution. For direct
comparison we chose an uniform scaling for both plots. The difference
in $(V-I)$ and $(U-I)$, respectively, between a 15 Gyr model isochrone
and the one for 13,~10,~7, and 5 Gyr (from bottom to top) is shown
here.}
\label{f:ageresol}
\end{figure}

In this paper we described our method of relative age dating in
globular cluster systems using integrated photometry and constrain its
suitability. The main emphasis is not the determination of ages with
only small uncertainties, as it has been done in various papers by
using spectroscopy for integrated light or globular clusters
(e.g. \cite{trager98}, \cite{kuntschner00}, Poggianti et al. 2001,
\cite{moore02}, \cite{puzia03}), but to introduce a method to identify
different age populations in large systems and to evaluate the
importance of the underlying star formation events. From the results
presented here we conclude that for large globular cluster sets
integrated photometry is a suitable tool to search for sub-populations
of globular clusters. The size of the young population appears
excessively large when compared to the maximum allowed young
population in the integrated light of these galaxies. Based on the
fact that we study all globular cluster systems in the center of
galaxies and assuming that intermediate clusters, formed during
merger, settle down in the new center of gravity, our data will be
biased toward these intermediate age objects. More observations
covering regions more distant to the center of the host galaxies are
necessary to overcome this bias effect. Yet, the bias alone is
unlikely to explain the fraction $>$80$\%$ found in NGC~5846 and
NGC~4365. Our example in Sect. 5.1, using an age of 10 Gyr for the old
population, resulting in fractions close to 40$\%$ of young clusters,
illustrates how the method is sensitive to this choice and/or proper
SSP calibration. We conclude that our idealized SSP models probably
still host uncertainties, especially do not take into account anything
but age and metallicity, e.g. $\alpha$/Fe varying abundance ratios
etc. might play a role and introduce additional scatter or systematic
error in the old population, driving our results to an interpretation
of overly large young populations. These effects will be investigated
in the future together with modelers. In particular the effect of
$\alpha$/Fe enhancement, as now incorporated into the models by
C. Maraston \& D. Thomas (\cite{thomas03},
\cite{maraston03}). Together with a modified calculation of the
secondary color (see Sect.\ref{s:second}) these are the next step for
improving our approach. For now the method is able to detect
intermediate age populations, roughly constrain their age but most
probably overestimates their importance. To lower the age uncertainty
in the colour-colour distributions the observations will be extended
into the blue wavelength region, e.g. by including the U-band. As
shown in Fig. \ref{f:ageresol} $(U-I)$$~vs.$$(V-K)$ color-color
diagrams will result in a considerably larger age resolution.

\vskip 0.5cm
{\it{Acknowledgments}} The authors would like to thank the
referee, A. Vazdekis for a careful review and many helpful remarks.


\begin{thebibliography}{}
\bibitem[Ashman \& Zepf~1992]{ashman92}Ashman, K.M., Zepf, S.E.~1992,
ApJ, 384, 50
\bibitem[Ashman et al.~1994]{ash94}Ashman, K.M., Bird, C.M., Zepf,
S.E.~1994, AJ, 108, 2348
\bibitem[Bennett et al.~2003]{bennett03}Bennett, C.L., Halpern, M.,
Hinshaw, G. et al.~2003, ApJ, accepted
\bibitem[Bertin \& Arnout~1996]{bertin96} Bertin, E., Arnout, S.~1996,
A\&AS, 117, 393B
\bibitem[Brodie 2002]{brodie02}Brodie, J.P.~2002, IAU Symposium Series, 207, 218
\bibitem[Bruzual \& Charlot~1993]{bruzual93}Bruzual, G.A. \&
Charlot, S.~1993, ApJ, 405, 538
\bibitem[Bruzual~2000]{bruzual00}Bruzual, A.G.~2000, private
communication 
\bibitem[Bruzual \& Charlot~2003]{bruzual03}Bruzual, G. \& Charlot, S.~2003, MNRAS, 344,
1000
\bibitem[Burgarella et al.~2001]{burgarella01}Burgarella, D., Kissler-Patig, M., Buat, V.~2001, AJ, 121,2647
\bibitem[Charlot \& Bruzual~1991]{charlot91}Charlot, S., Bruzual,
G.A.~1991, ApJ, 367, 126
\bibitem[Charlot et al.~1996]{charlot96}Charlot, S., Worthey, G., Bressan, A.~1996, AJ, 457, 625
\bibitem[de Grijs et al.~2003]{grijs03}de Grijs, R., Bastian, N.,
Lamers, H.J.G.L.M.~2003, ApJ, 583, L17
\bibitem[Elmegreen \& Efremov 1997]{elmegreen97}Elmegreen, B.G.,
Efremov, Y.N.~1997, ApJ, 480, 235 
\bibitem[Forbes et al. 1997]{forbes97}Forbes, D.A., Brodie, J.P., Grillmair, C.J.~1997, AJ, 113, 1652
\bibitem[Forbes \& Forte~2001]{forbes01}Forbes, D.A., Forte, J.C.~2001, MNRAS, 322,257
\bibitem[Gebhardt \& Kissler-Patig~1999]{gebhardt99}Gebhardt, K.\& Kissler-Patig,
M.~1999, AJ, 118, 1526
\bibitem[Geisler, Grebel \&  Minniti~2002 (eds.)]{geisler02}Geisler,  D., Grebel, E.K., Minniti, D. (eds.)~2002, ``Extragalactic Star Clusters'', IAU Symposium, 207
\bibitem[Goudfrooij 2001]{goudfrooij01}Goudfrooij, P., Alonso, M.V., Maraston, C.~et al.~2001, MNRAS, 328, 237 
\bibitem[Harris 1991]{harris91} Harris, W.E.~1991, ARAA, 29, 543
\bibitem[Harris 2003]{harris03}Harris, W.E.~2003, ``Extragalactic Globular Cluster Systems'', Springer Verlag, ed. Kissler-Patig, M, ESO Astrophysics Symposia, p317
\bibitem[Hempel et al.~2003]{hempel03}Hempel, M., Hilker, M.,
Kissler-Patig et al.~2003, A\&A, 405,487 (Paper III)
\bibitem[Infante \& Pritchet~1995]{infante95} Infante, L. \& Pritchet, C.J.~1995, AJ, 439, 565
\bibitem[Jord\'an et al.~2002]{jordan02}Jord\'an, A., C\^ot\'e, P.,
West, M.J. et al.~2002, ApJ, 576, L113
\bibitem[Kennicutt~1998]{kennicutt98b}Kennicutt, R.C.~1998, in ``Galaxies:Interactions and Induced Star Formation'', eds. Friedli, Martinet, Pfenniger, Springer Verlag 
\bibitem[Kissler-Patig~2000]{kissler00}Kissler-Patig, M.~2000, RvMA,13,13
\bibitem[Kiss\-ler-\-Patig et al.~2002]{kiss02a} Kissler-Patig, M.,
Brodie, P. B., \& Minniti, D.~2002, A\&A, 391, 441 (Paper I)
\bibitem[Kiss\-ler- Pa\-tig 2003 (ed.)]{kissler03b}Kissler-Patig, M.(ed.)~2003, ``Extragalactic Globular Cluster Systems'', Springer Verlag Berlin
\bibitem[Kundu \& Whitmore 2001a]{kundu01a}Kundu, A., Whitmore, B.C.~2001, AJ, 121,2950
\bibitem[Kundu \& Whitmore 2001b]{kundu01b}Kundu, A., Whitmore, B.C.~2001, AJ, 122,1251
\bibitem[Kuntschner~2000]{kuntschner00}Kuntschner, H.~2000, MNRAS, 315,184
\bibitem[Kurucz 1992]{kurucz92}Kurucz, R.L.~1992, IAU Symp.149, 225
\bibitem[Larsen~2001]{larsen01}Larsen, S.~2001, IAU Symposium Series, 207
\bibitem[Larsen et al.~2003]{larsen03}Larsen, S., Brodie, J.P., Beasley,
  M.A. et al.~2003, ApJ, 585, 767)
\bibitem[Lejeune et al.~1997]{lejeune97}Lejeune, Th.; Cuisinier, F.; Buser, R.~1997, A\&AS, 125, 229L
\bibitem[Maller et al.~2003]{maller03}Maller, A.H., McIntosh, D.H.,
Katz, N. et al.,~2003, ApJ, submitted (astro-ph/0304005) 
\bibitem[Maraston et al.~2001]{maraston01}Maratson, C., Greggio, L.,
Thomas, D.~2001, Ap\&SS, 276, 893
\bibitem[Maraston et al.~2003]{maraston03}Maraston, C., Greggio, L., Renzini, A. et al.,~2003, A\&A, 400, 823
\bibitem[Moore et al.~2002]{moore02}Moore, S.A.W., Lucey, J.R.,
Kuntschner, H. et al.~2002,MNRAS, 336, 382
\bibitem[Poggianti et al.~2001]{poggianti01}Poggianti, B.M, Bridges,
T., Carter, D. et al. ~2001, AJ, 563,118
\bibitem[Press et al. 1992]{press92}Press, W.E., Teukolsky, S.A., Vetterling, W.T. et al.~1992, ``Numerical Recipes in Fortran'', Cambridge University Press
\bibitem[Puzia et al.~1999]{puzia99}Puzia, T.H.,Kissler-Patig, M., Brodie, J.P.
 et al.~1999, AJ, 118,2743
\bibitem[Puzia et al.~2002]{puzia02} Puzia, T. H., Zepf, S. E.,
Kissler-Patig et al.~2002, A\&A, 391, 453 (Paper II)
\bibitem[Puzia et al.~2003]{puzia03}Puzia, T.H., Kissler-Patig, M.,
Thomas, D. et al. A\&A, in prep.
\bibitem[Renzini 1999]{renzini99a}Renzini, A.~1999, in ``The formation of galactic Gulges'', eds. Carollo, Ferguson \& Wyne, Cambridge University Press
\bibitem[Renzini \& Cimatti~1999]{renzini99b}Renzini, A., Cimatti, A.~1999, ASP Conference Proceedings, 193, 312
\bibitem[Roche et al.~1993]{roche93}Roche, N., Shanks, T., Metcalf, N. et al., ~1993, MNRAS, 263, 360
\bibitem[Schweizer et al.~1996]{schweizer96}Schweizer, F., Miller, B.W.,
Whitmore, B.C. et al.~1996, AJ, 112, 1839
\bibitem[Schweizer~2000]{schweizer00}Schweizer, F.~2000, Phil. Trans. R. Soc. Lond., Ser. A, 358,2063
\bibitem[Thomas et al.~2003]{thomas03}Thomas, D., Maraston, C.,\& Bender, R.,~2003, MNRAS, 339, 897 
\bibitem[Tinsley~1972]{tinsley72}Tinsley, B.M.~1972, ApJ, 179, 319
\bibitem[Toomre~1977]{toomre77}Toomre, A., in The Evolution of Galaxies
and Stellar Populations, ed. V.Trimble and R.B.Larson, ~1977, 401
\bibitem[Trager  et al.~1998]{trager98}Trager, S.C., Worthey, G.,
Faber, S.M.~ 1998, ApJS, 116,1 
\bibitem[Trager~2003]{trager03}Trager, S.C., in Carnegie
Observatories Astrophysics Series,~2003, 4, Cambridge University Press
\bibitem[Vazdekis et al.~1996]{vazdekis96}Vazdekis, A., Casuso, E.,
Peletier, R.F. et al.,~1996, ApJS, 106, 307
\bibitem[Vazdekis~1999]{vazdekis99}Vazdekis, A.~1999, ApJ, 513, 224
\bibitem[Worthey~1994]{worthey94} Worthey, G.~1994, ApJS, 95, 107
\end{thebibliography}
\end{document}